\documentclass[12pt,preprint]{aastex}

\newcommand*{\logg}{$\log~g$}
\newcommand*{\feh}{[Fe/H]}
\newcommand*{\afe}{[$\alpha$/Fe]}
\newcommand*{\kms}{km s$^{-1}$}

\newcommand*{\msun}{$M_\odot$}
\newcommand*{\teff}{$T_{\rm eff}$}
\newcommand*{\alp}{$\alpha$}

\usepackage{lscape}
\usepackage{epsfig}
\usepackage{apjfonts}
\usepackage{amsmath}
\usepackage{longtable}
\usepackage{leqno}
\usepackage{graphics}

\shorttitle{The SEGUE Stellar Parameter Pipeline. V.}
\shortauthors{Lee et al.}

\begin{document}

\title{The SEGUE Stellar Parameter Pipeline. V. Estimation of Alpha-Element Abundance Ratios From Low-Resolution SDSS/SEGUE Stellar Spectra}

\author{Young Sun Lee\altaffilmark{1}, Timothy C. Beers\altaffilmark{1}, Carlos Allende Prieto\altaffilmark{2,3,4},
David K. Lai\altaffilmark{5}, Constance M. Rockosi\altaffilmark{5}, Heather L. Morrison\altaffilmark{6}, Jennifer A. Johnson\altaffilmark{7},
Deokkeun An\altaffilmark{8}, Thirupathi Sivarani\altaffilmark{9}, Brian Yanny\altaffilmark{10}}

\altaffiltext{1}{Department of Physics and Astronomy and JINA (Joint Institute for Nuclear Astrophysics),
                 Michigan State University, East Lansing, MI 48824, USA; lee@pa.msu.edu, beers@pa.msu.edu}
\altaffiltext{2}{Instituto de Astrof\'{\i}sica de Canarias, E-38205 La Laguna, Tenerife, Spain}
\altaffiltext{3}{Departamento de Astrof\'{\i}sica, Universidad de La Laguna, E-38206 La Laguna, Tenerife, Spain}
\altaffiltext{4}{Mullard Space Science Laboratory, University College London, Holmbury St. Mary,
                 Dorking, RH5 6NT Surrey, UK; cap@mssl.ucl.ac.uk}
\altaffiltext{5}{UCO/Lick Observatory, Department of Astronomy and Astrophysics,
                 University of California, Santa Cruz, CA 95064, USA; david@ucolick.org, crockosi@ucolikc.org}
\altaffiltext{6}{Department of Astronomy, Case Western Reserve University,
                 Cleveland, OH 44106, USA; heather@vegemite.case.edu}
\altaffiltext{7}{Department of Astronomy, Ohio State University,
                 Columbus, OH 43210, USA; jaj@astronomy.ohio-state.edu}
\altaffiltext{8}{Department of Science Education, Ewha Womans University, Seoul 120-750, Republic of Korea; deokkeun@ewha.ac.kr}
\altaffiltext{9}{Indian Institute of Astrophysics, 2nd block Koramangala, Bangalore 560034, India; sivarani@iiap.res.in}
\altaffiltext{10}{Fermi National Accelerator Laboratory, Batavia, IL 60510, USA; yanny@fnal.gov}

\begin{abstract}

We present a method for the determination of \afe~ratios from low-resolution ($R
= 2000$) SDSS/SEGUE stellar spectra. By means of a star-by-star comparison with
degraded spectra from the ELODIE spectral library and with a set of moderately
high-resolution ($R = 15,000$) and medium-resolution ($R = 6000$) spectra of
SDSS/SEGUE stars, we demonstrate that we are able to measure \afe~from
SDSS/SEGUE spectra (with S/N $> 20/1$) to a precision of better than 0.1 dex,
for stars with atmospheric parameters in the range \teff~= [4500, 7000] K,
\logg~= [1.5, 5.0], and \feh~=~[$-$1.4, $+$0.3], over the range \afe~=~[$-$0.1, $+$0.6]. For stars
with [Fe/H] $< -1.4$, our method requires spectra with slightly higher
signal-to-noise to achieve this precision (S/N $> 25/1$). Over the full
temperature range considered, the lowest metallicity star for which a confident
estimate of \afe~can be obtained from our approach is [Fe/H] $\sim -2.5$;
preliminary tests indicate that a metallicity limit as low as [Fe/H] $\sim -3.0$
may apply to cooler stars. As a further validation of this approach, weighted
averages of \afe~obtained for SEGUE spectra of likely member stars of Galactic
globular clusters (M15, M13, and M71) and open clusters (NGC 2420, M67, and NGC
6791) exhibit good agreement with the values of \afe~from previous studies.
The results of the comparison with NGC 6791 imply that the
metallicity range for the method may extend to $\sim+$0.5.

\end{abstract}


\section{Introduction}

The chemical elements O, Mg, Si, Ca, and Ti, which are often referred to
as the ``\alp'' elements, are mainly produced in the interiors of
high-mass stars ($> 8$ \msun) during their main-sequence lifetime and
later explosive nucleosynthesis events. These elements, at least some
fraction of which are ejected into the interstellar medium (ISM) by
core-collapse supernovae (Timmes et al. 1995), enhance the next generations of
stars that form out of this polluted ISM. The abundances of the \alp-elements
for subsequent stellar generations thus increase very quickly, due to the
relatively short main-sequence lifetimes of massive stars (millions to tens of
millions of years). However, as Fe can be produced by the same process,
enhancements of the \afe\footnote[11]{This notation is defined by an average of
[Mg/Fe], [Si/Fe], [Ca/Fe], and [Ti/Fe] in many works, and it is similarly
applied in this paper.} ratios remain at roughly the same levels.
Substantially more Fe is generated by Type Ia supernovae (SNe Ia), 
whose progenitor lifetimes are much longer
(on the order of 0.1 Gyr to a few Gyr; Matteucci \& Recchi 2001).
Nucleosynthesis during these explosions is thought to produce a large
fraction of the iron-peak elements (Ni, Co, Fe, etc.; Nomoto et al. 1984).
Hence, an ISM polluted by material enriched with these iron-peak elements at
later times will naturally lead to stars that form with relatively low
\afe~ratios.

In the Milky Way, a typical metal-poor halo star exhibits abundance
ratios \afe~$\sim +0.4$ (e.g., McWilliam 1997). Thick-disk stars have
relatively higher abundance ratios (\afe~$\sim$ $+$0.3 to $+$0.4; Bensby et al.
2003, 2005; Reddy et al. 2006) than do thin-disk stars (\afe~$\sim$ 0.0 to
$+$0.1). Many stars in the dwarf spheroidal galaxies (dSphs) in the vicinity
of the Milky Way possess relatively lower \afe~compared to the Galactic
halo stars of the same metallicity (Shetrone et al. 2001, 2003;
Fulbright 2002; Tolstoy et al. 2003; Geisler et al. 2005), at least for
[Fe/H] $> -2.5$. It is thought that the low \afe~ratio in these systems
is due to a more inefficient star formation history, resulting in a greater
contribution of Fe from SNe Ia than occurred for the bulk of the halo
stars of the Milky Way (Unavane et al. 1996). Recent evidence has been
accumulating (e.g., Frebel et al. 2010, and references therein) that the
\afe~ratios for the lowest metallicity stars in dSphs (including the
majority of stars in the low-luminosity dSphs discovered by the Sloan 
Digital Sky Survey (SDSS; Zucker et al. 2006a, 2006b; Belokurov et al. 2006, 2007, 2008, 2010) are
quite similar to those observed for stars in the Galactic halo with
[Fe/H] $< -2.5$, strengthening the argument of Carollo et al. (2007,
2010) and Frebel et al. (2010) that dSphs may well have contributed to
(at least) the formation of the outer-halo component of the Galaxy.

As the evolution of chemical abundances in a given system is closely
related to the history of its star formation (e.g., Tinsley 1979),
abundance ratios such as \afe~can be used to explore these histories for
recognized components of the Galaxy and its satellites, in particular if
such measurements can be obtained for large numbers of member stars.

Past efforts to measure the abundance of \alp-elements have been mostly
based on high signal-to-noise raito (S/N), high-resolution ($R > 20,000$)
spectroscopy, since individual elemental lines can be identified
relatively easily and used to obtain the most accurate abundances.
However, with current generation telescopes and instrumentation, the
(typically) long exposure times required generally limit the samples of
targets to bright stars or small numbers of fainter stars, and it is not
(yet) feasible to obtain high-resolution spectra for the large numbers
of stars needed to systematically study the history of star formation
and chemical enrichment of the recognized Galactic components.

The above limitations have inspired investigations of techniques capable
of recovering estimates of abundances for \alp-elements using lower
resolution spectroscopy. For example, Kirby et al. (2008a) developed a
spectral matching technique based on medium-resolution ($R \sim 6000$)
spectra over the wavelength range 6300 -- 9100 \AA. This approach
simultaneously measures [Fe/H] and \afe, and has been validated by
comparing the derived atmospheric parameters with giants in seven
Galactic globular clusters (GCs) over a range of metallicity. The basic idea
of the method is to generate a grid of synthetic spectra spanning
various temperature, gravity, and metallicity ranges having different
\alp-abundances, and search this grid for the best-matching synthetic spectrum to a
given observed spectrum. The parameters of the best-matching synthetic
spectrum (i.e., the global minimum of $\chi^{2}$) are then adopted. In
their approach, Kirby et al. chose to fix \teff~and \logg~based on
measured photometry (and extrapolation of known relationships between
\teff~and \logg~for giant-branch stars). They report a precision for the
determination of [Fe/H] of about 0.1 dex, and as low as 0.05 dex for
\afe~for high-S/N spectra. They then applied this technique to derive
[Fe/H] for red giants in low-luminosity dSphs (Kirby et al. 2008b), and
similarly, to velocity-member stars in the classical Sculptor dwarf
galaxy (Kirby et al. 2009), and were able to estimate
\feh~and the abundances of Mg, Si, Ca, and Ti for stars in their sample
of dwarf galaxies.

As another example, one contemporary survey that obtains abundances of
the \alp-elements for large samples of stars is the Radial Velocity
Experiment (RAVE; Steinmetz et al. 2006), a spectroscopic survey for
(up to) a million stars in the southern sky. The spectral range explored
is 8410 -- 8795 \AA, at a resolving power $R = 7500$. Using an extensive grid
of synthetic spectra and a penalized $\chi^{2}$ matching technique,
reasonably accurate estimates of \teff, \logg, \feh, and somewhat less
accurate \afe~are obtained for the observed stars (Second Data Release;
Zwitter et al. 2008). RAVE is limited to bright stars ($9 < I < 12$), so
that the information gathered will be most useful for exploration of the
thin- and thick-disk populations.

The Sloan Extension for Galactic Understanding and Exploration (SEGUE;
Yanny et al. 2009) was one of three surveys that were executed as part
of the extension of the SDSS-II, which
comprised the sub-surveys Legacy, Supernova Survey, and SEGUE. The SEGUE
program was designed, in part, to obtain $ugriz$ imaging of some 3500 deg$^{2}$ of 
sky outside of the original SDSS-I footprint (Fukugita
et al. 1996; Gunn et al. 1998, 2006; York et al. 2000; Stoughton et al.
2002; Abazajian et al. 2003, 2004, 2005, 2009; Pier et al. 2003;
Adelman-McCarthy et al. 2006, 2007, 2008), and roughly 240,000
low-resolution ($R = 2000$) stellar spectra covering the wavelength range
3820 -- 9100 \AA. A one year dedicated survey devoted to obtaining such
spectra for fainter stars, SEGUE-II, executed as part of the ongoing
SDSS-III extension, has added an additional $\sim$140,000 stars. Stellar
spectra obtained during regular SDSS-I and SDSS-II operation add roughly
another 100,000 stars, for a total of about 500,000 stars with
potentially suitable spectra for further analysis.

These SDSS/SEGUE stellar spectra are processed through the SEGUE Stellar Parameter
Pipeline (SSPP; Lee et al. 2008a, 2008b, Allende Prieto et al. 2008, Smolinski
et al. 2010; Paper I, Paper II, Paper III, Paper IV respectively, hereafter), and the three
primary stellar parameters (\teff, \logg, and \feh) are reported for
most stars in the temperature range 4000 -- 10,000~K and with spectral
S/N ratios exceeding 10/1 (averaged over the the entire spectrum). The
SSPP estimates the atmospheric parameters via application of a number of
approaches, such as a minimum distance method (Allende Prieto et al.
2006), neural network analysis (Bailer-Jones 2000; Willemsen et al.
2005; Re Fiorentin et al. 2007), auto-correlation analysis (Beers et al.
1999), and a variety of line index calculations based on previous
calibrations with respect to known standard stars (Beers et al. 1999;
Cenarro et al. 2001a,b; Morrison et al. 2003). The SSPP adopts 6 primary
methods for estimation of $T_{\rm eff}$, 10 for log $g$, and 12 for
[Fe/H]. A series of additional procedures are used, especially for
[Fe/H], to obtain refined estimates of the final averages of the
multiple determinations for each parameter. We refer the interested reader to 
Paper I for more details on the SSPP. Paper I also performs a preliminary 
comparison with an average of two different high-resolution
spectroscopic analyses of over 100 SDSS/SEGUE stars, and claims that the SSPP is
able to determine $T_{\rm eff}$, log $g$, and [Fe/H] to precisions of 141~K,
0.23 dex, and 0.23 dex, respectively, after combining small systematic offsets
quadratically for stars with 4500~K $\leq T_{\rm eff} \leq$ 7500~K. 

Paper II shows how to select likely member stars of globular (M15, M13, and M2)
and open (NGC 2420 and M67) clusters, utilizing the SDSS photometry and
spectroscopy, and uses them to derive an overall metallicity of each cluster 
to validate the parameter-estimating ability of the SSPP, in
particular, [Fe/H]. The results of a comparison of the calculated overall
metallicities of the clusters to literature values suggest that the
metallicities derived by the SSPP have the typical external uncertainties of 0.13
dex, for a wide range of colors ($-0.3 \leq g - r \leq 1.3$), down to a
spectroscopic signal-to-noise of S/N = 10/1. Paper III describes how the
high-resolution spectra of the SDSS/SEGUE stars were analyzed and provides a
more thorough comparison with the SSPP parameters. The quoted errors of the
parameters derived by the SSPP in that paper are $\sigma(T_{\rm eff})$ = 130~K,
$\sigma(\log g)$ = 0.21 dex, and $\sigma([\rm Fe/\rm H])$ = 0.11 dex, which
slightly differ from those obtained by Paper I, although they share a common set
of high-resolution calibration observations. This is due to the fact that Paper
III derived the external uncertainties of the SSPP by only taking into account
the stars observed with the Hobby Eberly Telescope (HET; Ramsey et al. 1998),
while Paper I considered all available high-resolution spectra, including the
sample referred to as OTHERS in Paper III, which exhibits somewhat larger
scatter in its parameters when compared with those determined by the SSPP.

After adding five more GCs (M71, M3, M53, M92, and NGC 5053) and
three more open clusters (M35, NGC 2158, and NGC 6791), another round of
validation of the metallicity determined by the SSPP is performed in Paper IV,
which makes use of the same approach as in Paper II to select the likely member
stars of each cluster and to compute the overall metallicity. This paper confirms that
the typical scatter of the SSPP-derived [Fe/H] is in the order of $\sim$ 0.1
dex, as in Paper II. An appendix in Paper IV provides additional information on the
techniques used for the parameter averaging, and on changes made to the SSPP
since Paper I. In particular, the re-calibration of the metallicity estimated by
the {\tt NGS1} and {\tt NGS2} approaches results in improved metallicity
determinations for the metal-poor ([Fe/H] $< -2.0$) and metal-rich ([Fe/H] $\sim
0.0$) extremes (see Table 4 and the appendix of Paper IV).

Although the resolution of the SDSS/SEGUE spectra is too low to determine
abundances for individual \alp-elements, we have sought to estimate averaged
\afe~ratios by inspection of a specific wavelength range over which
\alp-element-sensitive features are found. The large dataset from SDSS/SEGUE
with available \afe~estimates is well suited for finding rare objects with low
or high \alp-element abundances, as well as for tracing the general trends
associated with different stellar populations.

In this paper, we describe a method for the determination of \afe~from
low-resolution SDSS/SEGUE spectra, and our efforts to validate this
technique, so that meaningful results can be obtained not only for stars
in the thin- and thick-disk components of the Galaxy, but also for stars
that are members of the halo populations. Our validation relies on
comparisons with degraded spectra for stars with known \afe~ratios taken
from the ELODIE spectral library (Moultaka et al. 2004), a
sub-sample of 91 of the roughly 350 SDSS/SEGUE stars for which we have
obtained high-resolution follow-up spectra to date, and likely members
of GCs and OCs with known \afe~ratios for which
the SEGUE spectra have been obtained. Section 2 describes our adopted
technique for estimation of \afe~for the SEGUE stellar spectra, while
the validation exercise is described in Section 3. Tests of the
reliability of our method, in particular its sensitivity to S/N and
degeneracies between surface gravity, metallicity, and the derived
\afe~ratios, are explored in Section 4. Section 5 presents a short
summary and conclusions.

\section{Methodology}
\subsection{Grid of Synthetic Spectra}

In order to obtain a fast, robust estimate of \afe~ratios for the SEGUE
spectra, we make use of a pre-existing grid of synthetic spectra. This
eliminates the need for generating synthetic spectra on the fly, while
simultaneously attempting to determine the primary atmospheric
parameters.

We have made use of Kurucz's NEWODF atmospheric models (Castelli \&
Kurucz 2003), with no enhancement of \alp-element abundances. These
models employ solar relative abundances from Grevesse \& Sauval (1998),
under the assumption of plane-parallel line-blanketed model structures
in one-dimensional local thermodynamical equilibrium (LTE), and include
H$_{2}$O opacity, an improved set of TiO lines, and no convective
overshoot (Castelli et al. 1997). The ready made models can
be downloaded from Kurucz's
Website\footnote[12]{http://kurucz.harvard.edu/grids.html}. We adopted 
these models to generate a finer grid (steps of 0.2 dex for \logg~and
0.2 dex for \feh) by linear interpolation between the wider model grids
(which have original steps of 0.5 dex).

Note that although there exist atmospheric models with \alp-element
enhancements, we have chosen not to use them, in order to avoid abrupt
changes in the model atmospheres introduced when interpolating to obtain a
finer grid in \afe. Employing a homogeneous set of model atmospheres
also ensures smooth changes in $\chi^{2}$ space as the parameters are
estimated. As a check on the impact of this choice, we compared some
synthetic spectra generated with \alp-enhanced models to those without
\alp-enhanced models, and found that the mean difference in the flux
between them is much less than 1$\%$, with a standard deviation smaller
than 1$\%$. Thus, the choice has minimal effect on the determination of
\afe.

We created synthetic spectra from this finer grid using the {\tt
turbospectrum} synthesis code (Alvarez \& Plez 1998), which uses the
treatment of line broadening described by Barklem \& O'Mara (1998),
along with the solar abundances of Asplund et al. (2005).
The sources of atomic lines used by {\tt turbospectrum} come largely
from the VALD database (Kupka et al. 1999). Line lists for the molecular
species CH, CN, OH, TiO, and CaH are provided by Plez (see Plez \& Cohen
2005, and B. Plez, private communication), while the lines of NH, MgH,
and the C$_{2}$ molecules are adopted from the Kurucz line
lists\footnote[13]{http://kurucz.harvard.edu/LINELISTS/LINESMOL/}.

When synthesizing the spectra we increase, by the same amount, the
abundances for the \alp-elements (O, Mg, Si, Ca, and Ti). A
micro-turbulence of 2 \kms~is adopted for all spectra, as at resolving
power $R = 2000$ the spectral features do not change measurably with
differently assumed values of the micro-turbulence (because the
micro-turbulence mostly influences the strong spectral lines). The
generated synthetic spectra have wavelength steps of 0.1 \AA. This
wavelength interval was chosen after considering the total computing
time required for carrying out the syntheses, and the fact that we found
little or no difference in the spectral line shapes for degraded
synthetic spectra formed by starting with steps smaller than 0.1 \AA.
For example, we found that the maximum difference in flux between a
spectrum with 0.1 \AA~and 0.005 \AA~steps, after smoothing to $R = 2000$,
is usually less than 2\%, and only increases to about 5\% for a cool 
metal-rich giant (e.g., \teff~= 4000 K, \logg~= 1.0, \feh~=
$+$0.4), located at the very edge of the parameter space of our grid.

\begin{figure}
\centering
\plotone{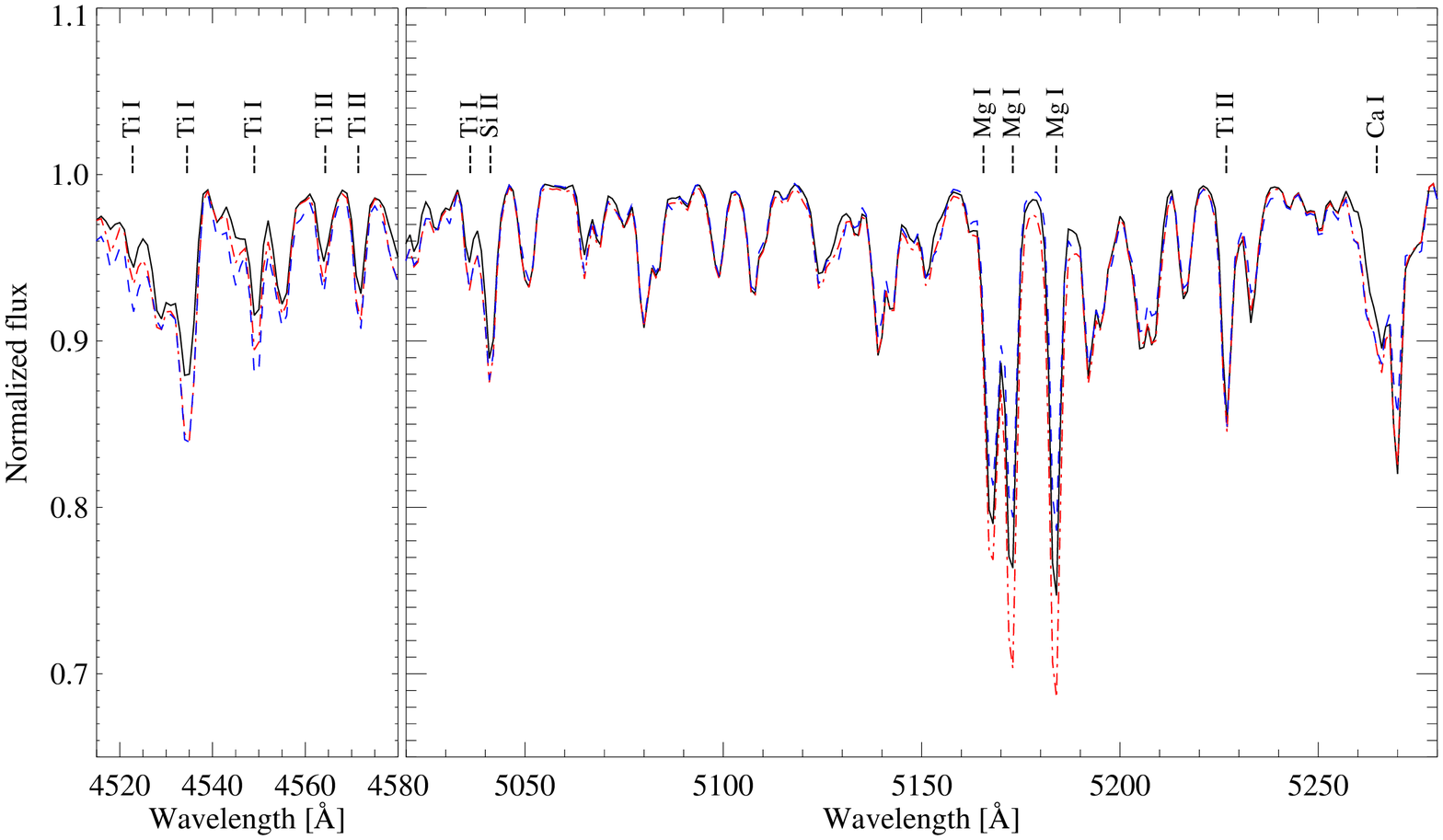}
\caption{Dominant \alp-element spectral features in the wavelength window
used to determine \afe. The black line shows a synthetic spectrum for a typical
dwarf of \teff~= 5500 K, \logg~= 4.4, \feh~$= -1.4$, and \afe~= 0.0. The red dash-dotted line
is the same, except \afe~= $+$0.4. The blue dashed line with \teff~= 5500 K, \logg~= 3.4,
\feh~$= -1.4$, and \afe~= $+$0.4 is also plotted. These spectra are degraded to the SDSS/SEGUE
resolution ($R = 2000$). It is clear that Mg and Ti contribute the most
to the determination of \afe, and furthermore, it is possible to easily
distinguish between the low and high \alp-abundance
cases.  However, it is also noticed that the three \ion{Mg}{1} and \ion{Ca}{1} line
strengths change with gravity.}
\label{fig:speclines}
\end{figure}

Each synthetic spectrum covers the wavelength range 4500 -- 5500 \AA.
This wavelength range was chosen because it contains a large set of
metallic lines, but avoids the CH $G$ band (4300 \AA), which is often a
strong feature in metal-poor stars, and the \ion{Ca}{2} K ($\sim$ 3933 \AA)
and H ($\sim$ 3968 \AA) lines, which become problematic for cool
metal-rich stars due to saturation of these lines. Most importantly,
this region includes prominent \ion{Mg}{1} and \ion{Ti}{1} and \ion{Ti}{2} lines (see Figure
\ref{fig:speclines}), which are very sensitive features for estimation
of the \alp-abundance.

The final grid covers 4000~K $\leq T_{\rm eff} \leq$ 8000~K in steps of
250~K, 0.0 $\leq \log~g \leq $ 5.0 in steps of 0.2 dex, and $-4.0 \leq
\rm [Fe/H] \leq +0.4$ in steps of 0.2 dex. The range in \afe~introduced
for the spectral synthesis covers $-0.1 \leq [\alpha/{\rm Fe}] \leq
+0.6$, in steps of 0.1 dex, at each node of $T_{\rm eff}$, log $g$, and
[Fe/H]. After creation of the full set of synthetic spectra, they are
degraded to SEGUE resolution ($R = 2000$) and re-sampled to 1 \AA~wide
linear pixels (during SSPP processing, the SEGUE spectra
are also linearly re-binned to 1 \AA~per pixel).

We do not attempt to determine the abundance of individual
\alp-elements; rather, we attempt to quantify their overall behavior. Hence,
our notation \afe~refers to an average of the abundance ratios for
individual \alp-elements, weighted by their line strengths in synthesized
spectra. As the dominant features in the spectral range we have selected
are Mg and Ti lines, these elements are certainly the primary
contributors to our determination of \afe, although Si and Ca may have
some influence, as seen in Figure \ref{fig:speclines}. 

It is clear that, if a star has very different abundances for the four
individual \alp-elements we consider (for the generation of synthetic spectra,
our assumption is that the abundances of the four elements vary in lockstep),
our measured \afe~may not correctly represent the overall content of the
\alp-elements, especially in cases of abnormally low (or high) Mg or Ti
abundances, since our technique relies heavily on the Mg and Ti lines.

Since O presents no strong detectable features in this wavelength range (at this
resolution) we exclude the O abundance when combining the \alp-element
abundances from the literature to compute the overall \alp-element
abundance, expressed as \afe, as we validate our measured \afe~with the
stars in other external sources such as the ELODIE spectral
library.

Owing to the different line strengths of each \alp-element in the
spectral window used to estimate \afe, we place different weights on
each element, and then compute the weighted mean of \afe~and its standard
deviation for the literature values in order to validate our
method of determining \afe:

\begin{equation}
\displaystyle
  <x> = \frac{\sum^{n}_{i=1}w_{i}x_{i}}{\sum^{n}_{i=1}w_{i}},
\label{eqn:mean}
\end{equation}

\begin{equation}
\displaystyle
  \sigma^{2} = \frac{\sum^{n}_{i=1}w_{i}(x_{i}-<x>)^{2}}{\sum^{n}_{i=1}w_{i}},
\label{eqn:rms}
\end{equation}

\noindent where $x_{i}$ is replaced by [Mg/Fe], [Ti/Fe], [Ca/Fe], or [Si/Fe], with weighting
factors $w_{i}$ = 5, 3, 1, or 1, respectively; $<x>$ is \afe. This
weighting system was determined after examination of line strengths
(equivalent widths, EWs) of individual \alp-elements shown in Figure
\ref{fig:speclines}. From the grid of synthetic spectra, we
calculated the EW of each line listed in Figure
\ref{fig:speclines}, summed up all EWs for all lines for each element,
and determined an averaged ratio of the summed EWs of each element to the total
EWs of all four elements. Through these computations we obtained
ratios of 0.50 for Mg, 0.33 for Ti, 0.08 for Ca, and 0.09 for Si
(rounded to 0.5 for Mg, 0.3 for Ti, 0.1 for Ca, and 0.1 for Si). For
elements without reported abundances in the literature for a given
star, zero weight is assigned ($w_{i}$ = 0). When less than four
elements are reported, the individual weights change accordingly, so
that the sum of the weights always adds to unity.

\subsection{Preprocessing Target and Synthetic Spectra}

The initial steps for the determination of \afe~for the SEGUE spectra
are to transform the wavelength scale to an air-based (rather than the
original SDSS vacuum-based) scale, and to shift the spectrum to the
rest frame using the radial velocity delivered by the SSPP. Following
these steps, the spectrum is linearly re-binned to 1 \AA~pixels over the
wavelength range 4500 -- 5500 \AA.

The spectrum under consideration is then normalized by dividing its
reported flux by its pseudo-continuum shape. The pseudo-continuum over
the 4500 -- 5500 \AA~range is obtained by carrying out an iterative
procedure that rejects points lying more than 1$\sigma$ below and
4$\sigma$ above the fitted function, obtained from a 9th-order
polynomial. Although we have a ``perfect'' continuum available for a
given synthetic spectrum, the synthetic spectra used to match with the
observed spectra are normalized in exactly the same fashion as for the
observed spectra, over the same wavelength range, and with
the same pixel size. Application of the same continuum routine ensures the
same magnitude of line-strength suppression in both spectra.

\subsection{Determination of \afe}

Following the above steps, we then search the grid of synthetic spectra
for the best-fitting model parameters. In this approach, we seek to
minimize the distance between the normalized target flux, $\emph{T}$,
and the normalized synthetic flux, $\emph{S}$, as functions of $T_{\rm
eff}$, log $g$, [Fe/H], and \afe, using a reduced $\chi^{2}$ criterion:

\begin{equation}
\chi^{2} = \frac{1}{m-n}\sum_{i=1}^{m}(T_i - S_i)^{2} /\sigma_i^{2},
\label{eqn:chi}
\end{equation}

\noindent where $\sigma_i$ is the error in flux in the $i$th pixel, $m$ is the number
of data points, and $n$ is the degrees of freedom. The SDSS/SEGUE
spectra provide the error in the flux associated with each pixel.

The parameter search over the reduced $\chi^{2}$ space is performed by
the IDL function minimization routine AMOEBA, which uses a downhill
simplex method (Nelder $\&$ Mead 1965). In this search process, we fix
\teff~to the value determined (previously) by the SSPP, and change
\logg, \feh, and \afe~simultaneously to minimize $\chi^{2}$, rather than
vary all four parameters at once. Since temperature has (by far) the
largest effect on the appearance of a stellar spectrum over the
wavelength range we consider, holding it constant permits the more
subtle variations associated with the other parameters to be explored.
Note that even though we adopt for convenience the notation ``SSPP''
parameters in the figures and tables in this paper, we make use of the
parameters (\logg, \feh, and \afe) determined by this method to compare
and validate, except \teff, which comes directly from the SSPP. It should be 
also kept in mind that as this technique shares the same grid of 
synthetic spectra as in {\tt NGS2}, one of the parameter searching techniques 
in the SSPP, we apply the same correction function (Equation (A2) in Paper IV) which 
was derived by re-calibrating the metallicity of the method to [Fe/H] determined 
by this method.

\begin{figure}
\centering
\plottwo{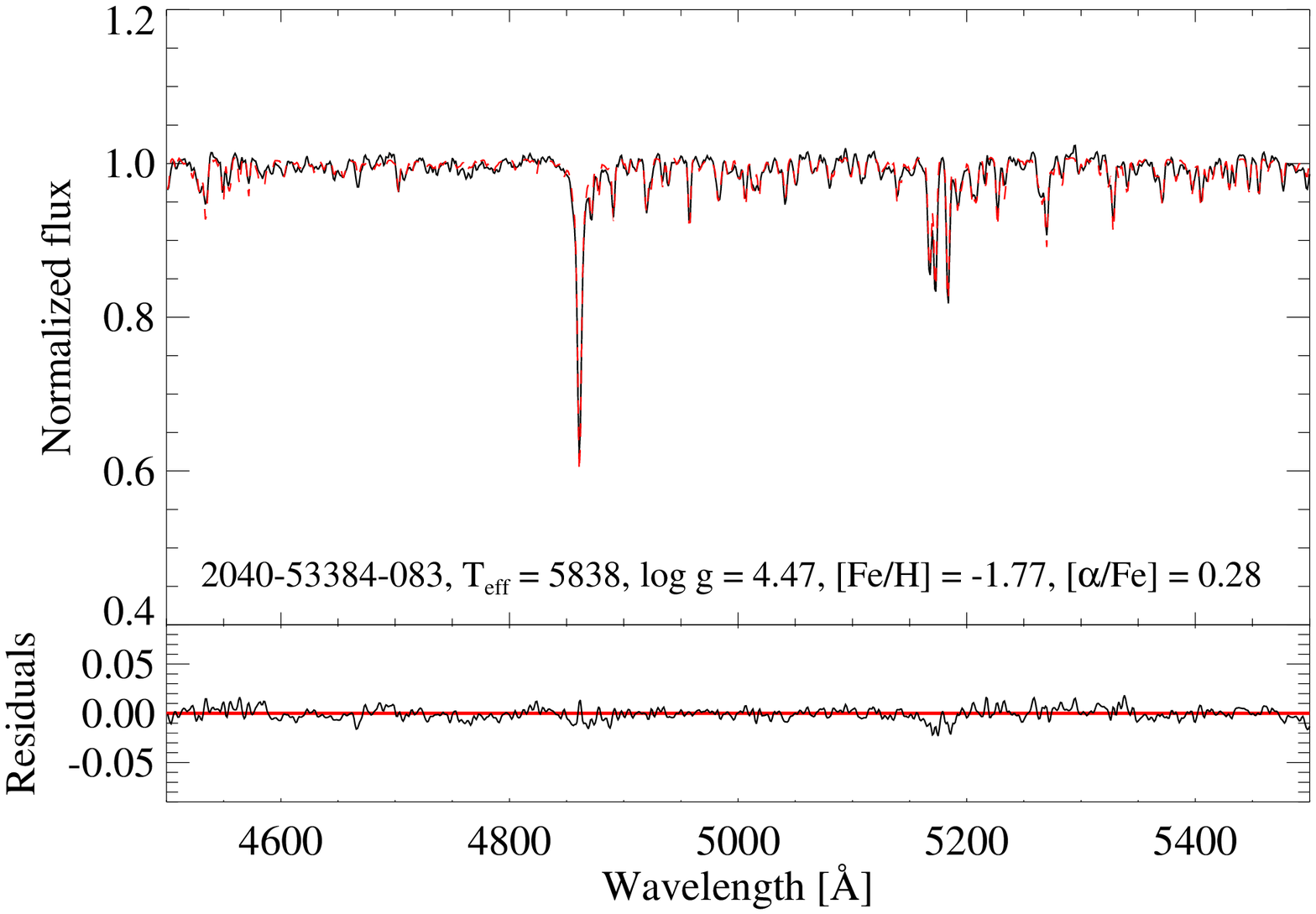}{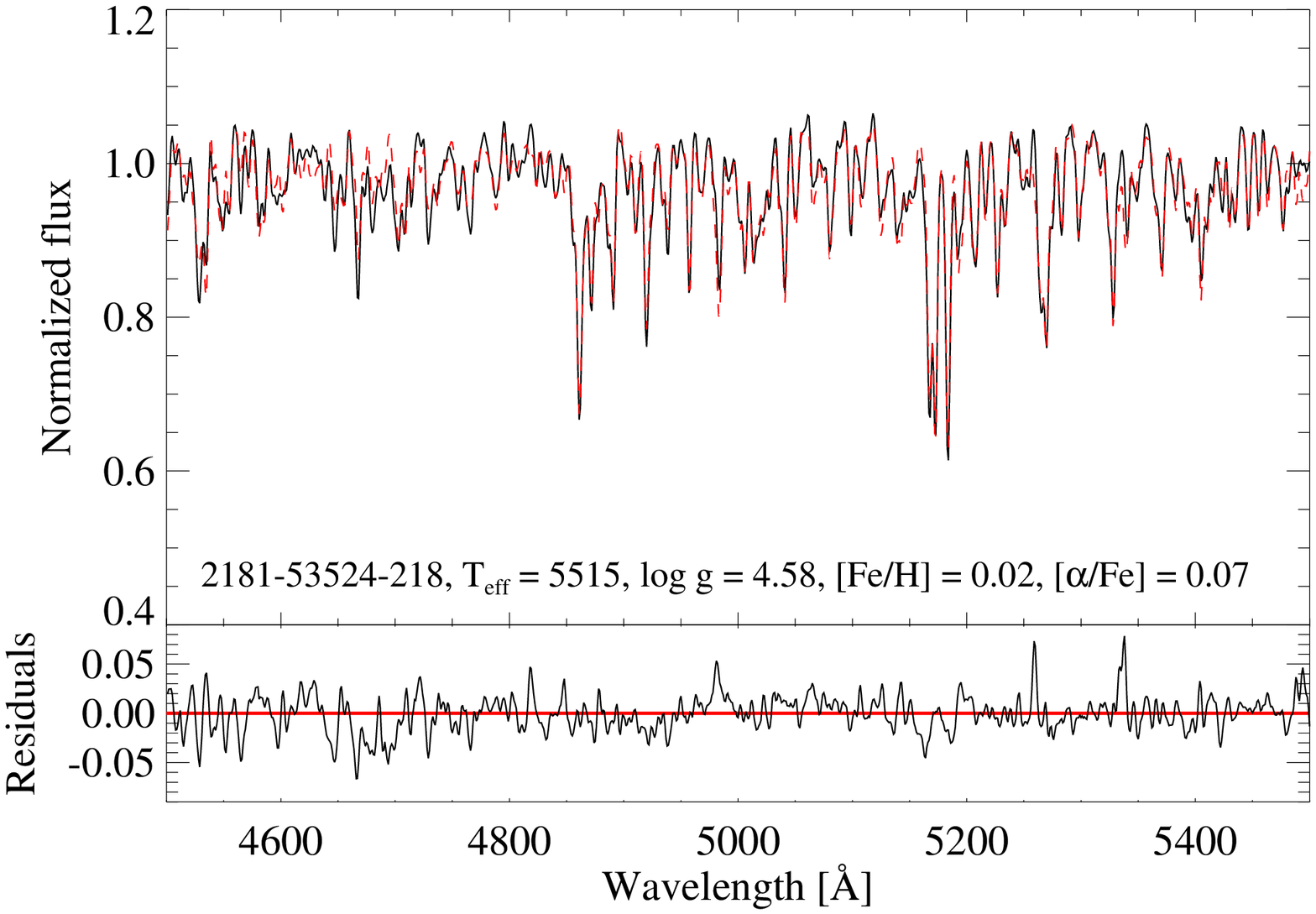}
\caption{Two examples of spectral matching. The left-hand panel
shows a metal-poor dwarf, while the right-hand panel shows a metal-rich dwarf.
The black line is the observed spectrum, while the red dashed line is
the best-matching synthetic spectrum generated with the parameters listed on each plot,
as determined by our methodology (with the exception of \teff; see the text). Residuals
between the observed and the synthetic spectrum are shown at the
bottom of each panel. Note that the largest residuals are no larger than 2\% for
the metal-poor star and 5\% for the metal-rich star.}
\label{fig:mpmr}
\end{figure}

Figure \ref{fig:mpmr} provides examples of the results of our spectral
fitting method. The left-hand panel shows an observed spectrum (black
line) for a mildly \alp-enhanced metal-poor dwarf, while the right-hand
panel corresponds to a slightly cooler, low \alp-abundance, metal-rich
dwarf. The red dashed line is the synthetic spectrum generated with the
parameters listed in each panel, while \logg, \feh, and \afe~are
determined by the method described above (\teff~is delivered by the
SSPP). From inspection, one can see that an excellent match between the
synthetic and observed spectra is achieved for these two stars. The
distribution of residuals shown at the bottom of each panel indicates
that the largest deviations for individual features are no more than a
few percent in the case of the metal-poor dwarf, and no more than five
percent for the metal-rich dwarf.

Errors associated with the determination of \afe~are estimated as
follows. We begin with the reported uncertainty associated with each
pixel, as delivered from the SDSS pipelines, and construct 10 different
realizations of the noise flux, assuming that the uncertainty is a
1$\sigma$ error of a Gaussian distribution. Then, after adding (or
subtracting) these noise fluxes to the observed flux, we follow the
method described above to determine \afe. Following this procedure, we
have 10 different estimates of \afe; the random error associated with
the measured \afe~is taken to be the standard deviation of these
different estimates. A more detailed discussion of
uncertainties in the \afe~determination is addressed below.

\section{Validation of the Method}

Once a technique to determine a physical quantity (\afe~in this
case) is constructed, it must be calibrated and validated with external
measurements. Rather than comparing the overall properties of a sample
of stars, it is preferable to compare star-by-star, ideally against
different sources, in order to quantify possible systematic offsets (and
optionally remove them), as well as to determine the likely random
errors associated with the estimate. For these purposes we employ the
ELODIE spectral library (Moultaka et al. 2004), as well as a set
of SDSS/SEGUE stars observed at moderately high dispersion ($R \sim 15,000$) 
with the High Resolution Spectrograph (HRS; Tull 1998) on the
HET (Ramsey et al. 1998) and at medium
resolution ($R = 6000$) with the Echellete Spectrograph and Imager (ESI;
Sheinis et al. 2002) on the Keck telescope. For convenience, in this
paper we refer to the spectra of SDSS/SEGUE stars observed with the HET
as ``HET'' spectra (data, or stars); those observed on the Keck are referred to as
``ESI'' spectra (data, or stars). In addition to the above samples, \afe~ratios for
likely member stars of the Galactic GCs (M15, M13, and M71)
and OCs (NGC~2420, M67, and NGC 6791) having SDSS/SEGUE spectra are
compared with published average values of \afe~for each cluster.

\subsection{Comparison with the ELODIE Spectral Library}

Spectra in the publicly available ELODIE library (we used version
ELODIE.3.1; Moultaka et al. 2004), were obtained with the ELODIE 
spectrograph at the Observatoire de Haute-Provence 1.93 m
telescope, and cover the spectral region 4000 -- 6800 \AA. As the higher
resolution spectra (with resolving power $R =  42,000$) are already
normalized to a pseudo-continuum, we employed the spectra with $R = 10,000$. 
Most of the spectra have quite high S/N ratios ($>$ 100 --
200/\AA), and are accompanied by estimated stellar parameters (not
including \afe) from the literature. Each spectrum (and parameter
estimate) has a quality flag assigned to it, ranging from 0 to 4, with 4
being best. In our comparison exercise, we only select stars with 4000~K
$\leq$ \teff~$\leq$ 8,000~K with a quality flag $\geq$ 1 for the spectra
and all of the parameters. The temperature range corresponds to that
covered by the grid of the synthetic spectra employed in our approach.

Although the spectra come with \teff, \logg, and \feh~estimates from
other studies and from the ELODIE library's own measurement, the
catalog does not supply \afe~measurements. The ``known'' \alp-element
abundances for the selected spectra are obtained from a literature
search, using the
VizieR\footnote[14]{http://webviz.u-strasbg.fr/viz-bin/VizieR} database.
The elemental abundances for these stars are based on high-resolution
analyses from various studies. Even though there may exist systematics
in the compiled \alp-element abundances between individual studies, we
do not attempt to adjust for any potential offsets from study to study.
As mentioned by Venn et al. (2004, and references therein), the
study-to-study and model-to-model variations in these abundances could
potentially be as large as 0.1 -- 0.2 dex.

From Equations (\ref{eqn:mean}) and
(\ref{eqn:rms}) we obtain a weighted average (and standard deviation) of
the available \alp-element measurements among Mg, Ti, Ca, and Si for a
given star, and represent them as \afe~and its error in our subsequent
analysis. Because some of the reported abundances among the four
elements are not accompanied by errors in the source papers, to report
the uncertainties in a consistent manner we conservatively estimate the
uncertainty in the reported \afe~from the standard deviation of the
different reported values of the four individual abundances. Recall that
we do not consider the O abundance, because the wavelength region chosen
for our analysis does not include any strong O lines.

We attempt to only use stars for which the abundances of at least three
\alp-elements are available; the Mg abundance must exist among the
three, and the abundance of the \alp-elements should also lie within the
range of our grid, $-0.1 \leq$ \afe~$\leq +0.6$. However, as we find that
most of metal-poor stars (\feh~$< -1.5$) in the ELODIE spectra
have only a reported Mg abundance, and they are valuable sources of
validation for our technique, we adopt [Mg/Fe] as an estimate of \afe~in
order to include them in our comparisons (\afe~uncertainties are not
reported for those stars in Table \ref{tab:elodie}).

\begin{figure}
\centering
\plotone{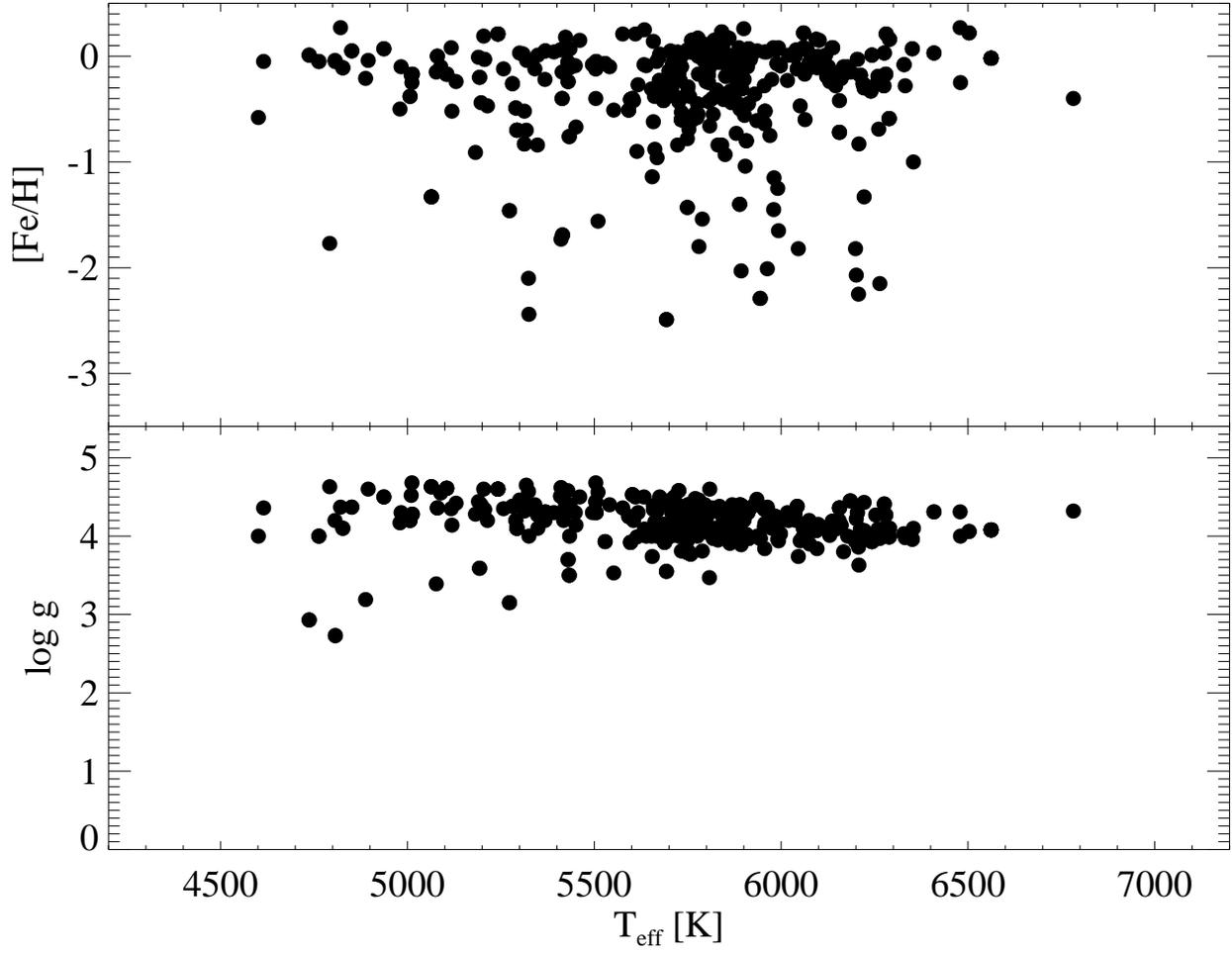}
\caption{Distribution of the stars in the ELODIE sample over 
\teff, \logg, and \feh~parameter space.}
\label{fig:elodiedist}
\end{figure}

Following these procedures, we have compiled a list of 293 unique stars with
good quality spectra (generally this corresponds to S/N $> 100/$\AA), with
atmospheric parameters in the range of our synthetic grid, and with available
information from which the determination of \afe~can be made. There are multiple
spectral observations made for some stars in the ELODIE spectra; we analyze the
individual spectra in our comparisons as they are good sources of consistency
checks on the derived parameters. Because of the inclusion of multiple spectra
for some stars, the total number of spectra considered in the comparison (425)
is much larger than the number of unique stars. Figure \ref{fig:elodiedist}
shows the parameter space that the ELODIE sample spans. Clearly, we see there
are not many cool metal-poor or metal-rich giants. The lack of metal-poor
giants is partially filled by the high-resolution sample discussed below.

The spectra of the stars from the sample meeting our desired
criteria are processed in the same way as the synthetic spectra used for
spectral matching, after degrading them to $ R = 2000$ using a Gaussian
kernel and re-sampling to 1 \AA~per pixel in the spectral range 4500 --
5500 \AA; \afe~for the selected stars is then estimated. It should be
noted that, due to the limited spectral range of the ELODIE
spectra compared with that of the SEGUE spectra, it is not possible to
process these spectra through the SSPP to obtain the stellar parameters,
in particular \teff, which is necessary to estimate \afe~with our
method. Therefore, we adopt \teff~from the literature value reported in
the ELODIE spectra, and hold it fixed during the $\chi^{2}$
minimization scheme.

Concerning error estimates for \afe, as the ELODIE spectra do not
provide explicit errors in the flux for each pixel of a given spectrum,
we conservatively assume a 1$\%$ error (which is a reasonable assumption
as S/N is 100/1 -- 200/1 for all spectra), and attempt to determine
measured errors in \afe~following the procedure outlined in Section 2.3.
Table \ref{tab:elodie} lists all the stars with the parameters adopted
in our comparison, including our measured parameters.

\begin{figure}
\centering
\plotone{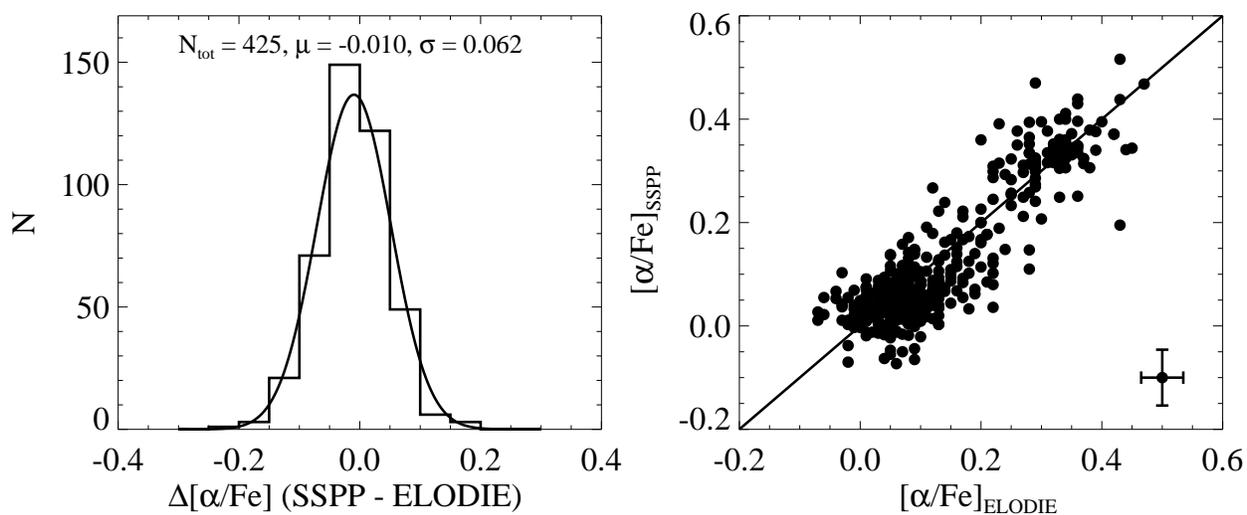}
\caption{Comparisons of \afe~obtained from
our estimates (SSPP) with 425 selected spectra from the ELODIE sample. 
The left-hand panel shows a Gaussian fit to the residuals
between \afe~for our method and values calculated from the literature,
while the right-hand panel is a run of our determination of \afe~against
the values for the ELODIE spectra. The black line in the
right-hand panel is the perfect correlation line. In this panel,
individual error bars are suppressed for clarity. Instead, a typical
error bar is shown in the lower-right corner.}
\label{fig:elodieafe}
\end{figure}

Figure \ref{fig:elodieafe} illustrates the results of our comparisons with the
selected ELODIE stars. ``ELODIE'' indicates the value of
\afe~obtained from the literature, while ``SSPP'' denotes our determination.
The left-hand diagram of the figure is a Gaussian fit to the residuals
between our values and those from the literature; an overall offset of
$-$0.010 dex is found, along with a scatter of 0.062 dex. The
right-hand panel plots our measurements against the literature values, and a
one-to-one correspondence line. There is no obvious deviation in our
determination along the perfect correlation line. For the sake of clarity, error
bars for each star in Figure \ref{fig:elodieafe} are suppressed, and a typical
total error bar in our measured \afe~and the literature values is denoted in the
lower right corner of the right-hand panel of the figure.

\begin{figure}
\centering
\plotone{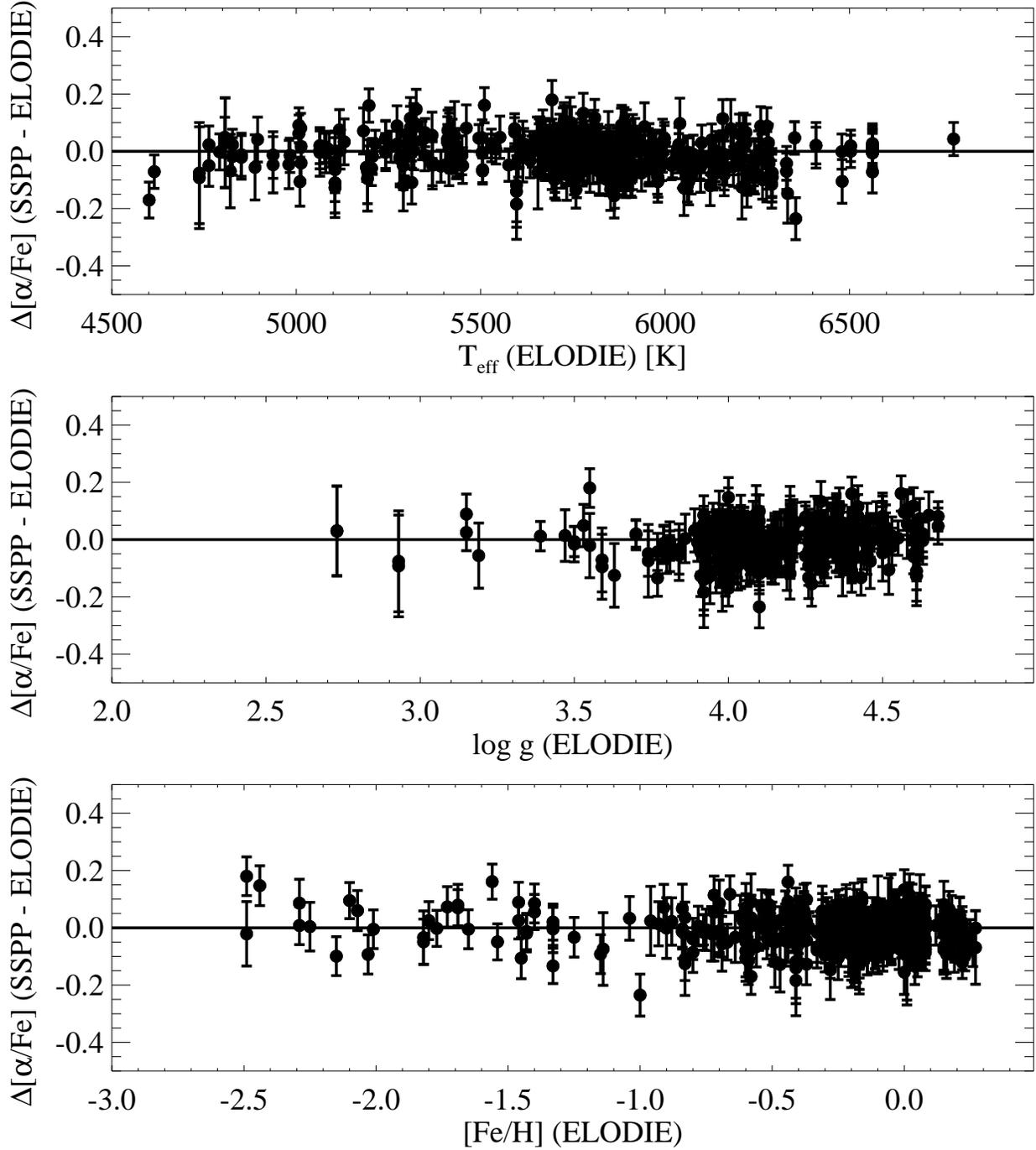}
\caption{Variations in \afe~as functions of \teff, \logg, and \feh, from
upper to lower panels, respectively, for the ELODIE spectra.
Quadratically added vertical error bars are shown.
There appears to be little or no covariance in the determined \afe~along with various
ranges for each parameter.}
\label{fig:elodiecor}
\end{figure}

Figure \ref{fig:elodiecor} displays residuals in \afe~between our
derived values and the ELODIE-derived values as a function of \teff, \logg,
and \feh~from upper to lower panels. Although it appears that there
is a small downward trend at low \teff~($<$ 4800 K) and our estimated
\afe~tends to be higher at \feh~$< -2.3$, the deviations seen
at such low \teff~and \feh~are mostly consistent with zero within the measured errors,
suggesting that there is no obvious correlation along with each parameter.

As the uncertainty in our measured \afe~consists of both systematic and random
errors, and the abundances that we employ from the literature (based on the
high-resolution analysis) also have uncertainties, we quantify the total error
in our measurement of \afe~by the following procedure. Let $\sigma_{\rm
g}$ be the rms scatter from a Gaussian fit to the residuals between our
measurements and the literature values of \afe, and $\sigma_{\rm HR}$ be
the error from the literature estimate calculated from Equation
(\ref{eqn:rms}). Then the systematic error ($\sigma_{\rm sys}$) in our measured \afe~is derived by 

\begin{equation}
\sigma_{\rm sys}^{2} = \sigma_{\rm g}^{2} - \sigma_{\rm HR}^{2} - \sigma_{\rm SSPP}^{2},
\label{eqn:sys}
\end{equation}

\noindent where $\sigma_{\rm SSPP}$ is the random error, simply taken to be the internal uncertainty of our
technique, estimated by following the procedure outlined in Section 2.3. In this
equation, $\sigma_{\rm HR}$ and $\sigma_{\rm SSPP}$ are an individual value for
each target, whereas a value of $\sigma_{\rm g}$ from the full sample is used.
That is, $\sigma_{\rm g}$ is fixed for all spectra, while $\sigma_{\rm HR}$ and
$\sigma_{\rm SSPP}$ change for each star. If $\sigma_{\rm HR}$ is not available
for a star, then we plug into the equation an average of $\sigma_{\rm HR}$ from
the sample. Using this relation, we compute an overall systematic error in our
measured \afe~by means of taking average of the sample, and define it as
$<\sigma_{\rm sys}>$. This $<\sigma_{\rm sys}>$ is applied to individual spectra
to yield the total error in our measurement of \afe~for each object by the
following equation:

\begin{equation}
 \sigma_{\rm tot}^{2} = <\sigma_{\rm sys}>^{2} +~\sigma_{\rm SSPP}^{2}.
\label{eqn:tot}
\end{equation}

In the equations above, the largest contribution to the total error
($\sigma_{\rm tot}$) comes from the scatter between our measured values
and the literature ones ($\sigma_{\rm g}$). However, if the noise in a
given spectrum dominates, the random error of ($\sigma_{\rm SSPP}$)
contributes more to the total error; this is discussed further in
Section 4.1.

The error bars in Figure \ref{fig:elodiecor} are obtained from the quadratic
addition of our measured total error to the literature error. Note that, as
mentioned earlier, there are some stars without properly measured errors. For
those stars, we adopt an average
\afe~uncertainty based on all stars with available error estimates. We also
notice from Figure \ref{fig:elodieafe} that the mean offset ($-0.010$
dex) is negligible, so we do not take it into account in our total error
calculations. We obtain 0.054 dex as the typical total error in our
\afe~estimates for the ELODIE spectra (obtained from a simple average of the
total errors for all stars considered in the figure).
The systematic error ($<\sigma_{\rm sys}>$) for the ELODIE spectra is 0.044 dex.
The rms scatter between this method and the literature is 0.092 dex for
\logg~and 0.122 dex for \feh, suggesting that the technique is robust for
the estimation of other parameters as well as for simultaneously determining \afe. 

All error bars in \afe$_{\rm SSPP}$ in the figures and the quoted total
errors of our estimated \afe~in the tables in this paper are calculated
from Equations (\ref{eqn:sys}) and (\ref{eqn:tot}) above.

\subsection{Comparison with SDSS/SEGUE High-resolution Spectra}

The optimal test of our ability to predict \afe~ratios from the SEGUE spectra comes
from a comparison with that obtained from analysis of higher-resolution spectra
observed for the same stars. Fortunately, a large number of suitable spectra
were obtained during the course of our efforts to calibrate atmospheric
parameters for the SSPP. We discuss this comparison below. Detailed information 
on these high-resolution spectra and their analysis
can be found in Papers III and IV, which validate (calibrate) the stellar parameters
determined by the SSPP by comparison with the high-dispersion spectra.

The general procedure for estimating \afe~for the HET spectra is
very similar to the analysis described in Paper III. The only differences are
that a slightly smaller wavelength window is used, and the inclusion of O, Mg,
Si, Ca and Ti elements, all with the same level of enhancement. The
determination of \afe~is performed by minimizing $\chi^{2}$ between the HET
spectra and the model fluxes, as in Paper III, but holding \teff~constant at the
value determined in Paper III, and fitting \logg, \feh, and
\afe~simultaneously. In this process, the internal error in each
parameter is derived from the square root of the diagonal elements of
the covariance matrix. The estimated \feh~and \logg~from this
minimization are consistent with those found in Paper III (the rms
scatter between them is 0.05 dex for \feh~and 0.16 dex for \logg,
respectively). 

We set the zero point for the \afe~values by forcing \afe~= 0 at 
\feh~= 0, which requires applying an offset of 0.13 dex over the entire 
metallicity range to the \afe~values determined directly from the analysis. 
This zero-point offset is independent of any other parameter and likely 
results from the usual approximations, in particular the use of a one-dimensional 
atmospheric structure in Local Thermodynamic Equilibrium (LTE). Allende Prieto
(2006) addressed the offsets in \afe~at \feh~= 0 in several surveys, 
and whether they might signal that the Sun is somewhat chemically
peculiar (see also Gustafsson 1998). His conclusion was that when solar 
analogs were examined (that is, when the effective temperature, surface 
gravity, and overall metal content of the stars were restricted to a narrow 
interval around the solar parameters) the offsets went away. This is the primary 
motivation for removing the 0.13 dex offset in our case.

From this analysis, we have obtained a total of 73 HET stars with well-measured
parameters, including \afe, in the range of our grid ($-0.1 \leq$ \afe~$\leq
+0.6$).

For the ESI spectra, Lai et al. (2009) describe the methods for determining the
stellar parameters, including the \alp-element abundances; we refer interested
readers to that paper. The abundances of Mg, Ti, and Ca are available for the
ESI stars, so we place zero weight on Si and the same weighting factors (5, 3,
and 1) on the others as before. Following Equations (\ref{eqn:mean}) and
(\ref{eqn:rms}), we calculate \afe~and its standard deviation for each star and
collect a total of 18 stars with available parameters within the boundaries of
our grid of synthetic spectra. The ESI data are mostly metal-poor giants, and
provide a useful sample to validate our technique for very metal-poor stars with
low surface gravity, filling the hole in the parameter space explored by the
ELODIE sample. Taken together with the HET spectra, we have a total of 91 stars
with which to compare. Table \ref{tab:hires} lists these stars, along with
the parameters and their associated errors from the high-resolution analysis and
from our measurements.

\begin{figure}
\centering
\plotone{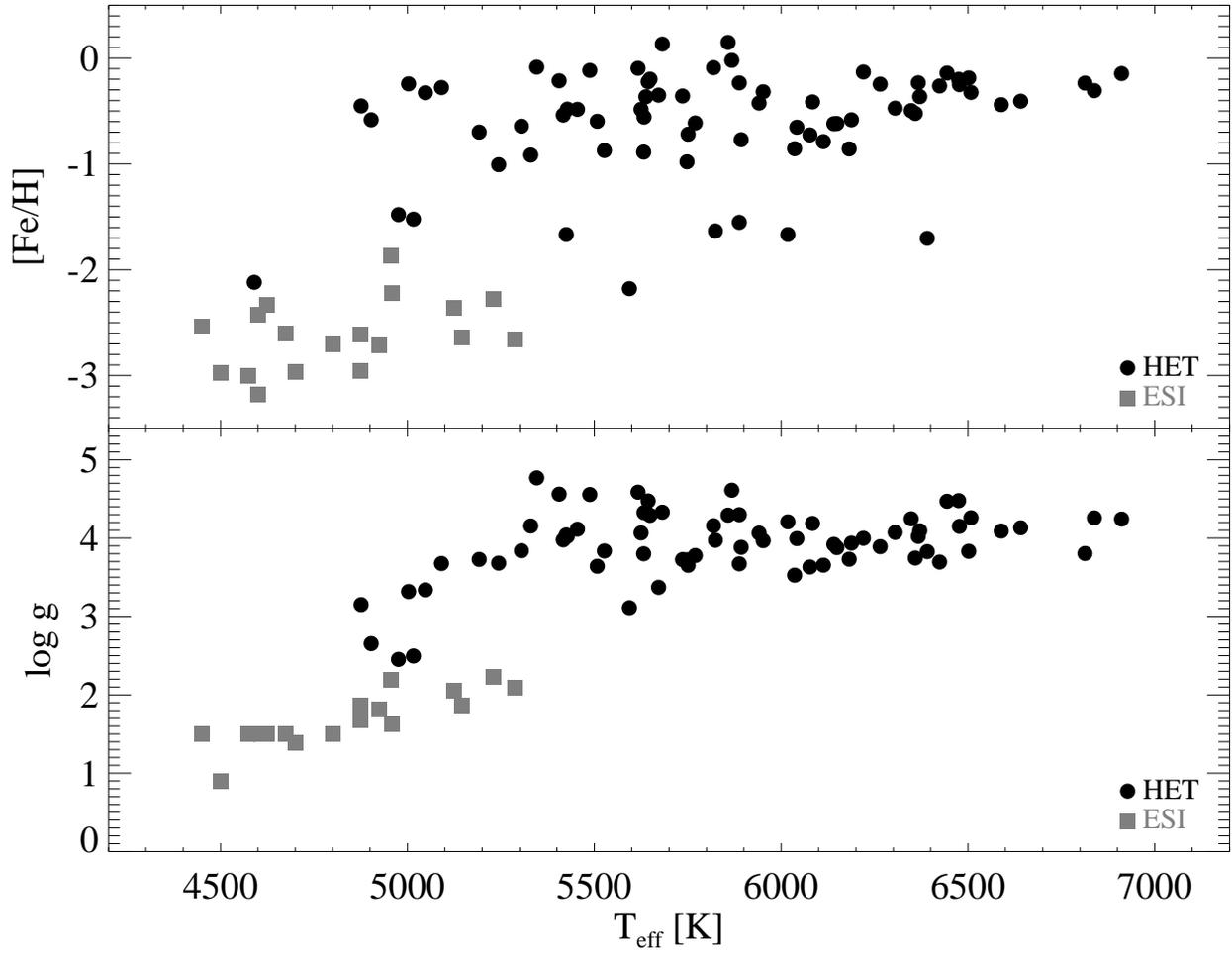}
\caption{Distribution of the stars with the high-resolution spectra over \teff, \logg, and \feh~
parameter space. The black dots are the HET spectra, while the gray squares are the ESI spectra.}
\label{fig:hiresdist}
\end{figure}

Figure \ref{fig:hiresdist} displays the range of the parameters for the
high-resolution sample. We note that the ESI data (gray squares) nicely cover
the gap in the cool metal-poor giant region, missing in Figure
\ref{fig:elodiedist}, but there remains a lack of cool metal-rich giants.
We attempt to address the defect in this regime by comparison
with a metal-rich open cluster, NGC 6791, in the following section.

\begin{figure}
\centering
\plotone{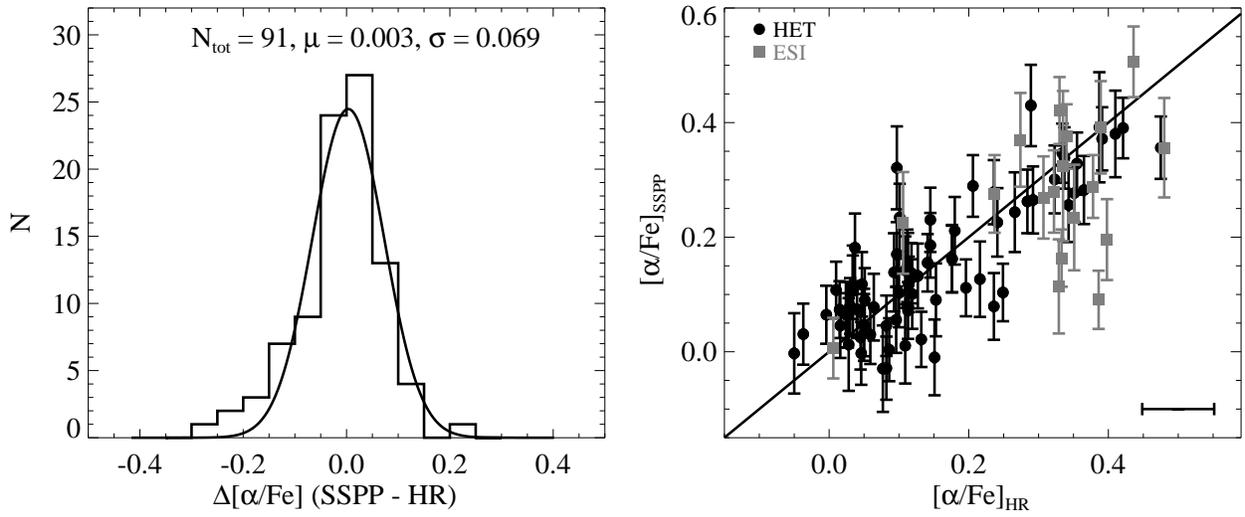}
\caption{Comparisons of \afe~from
our estimates (SSPP) for 91 stars having either HET (black dots) or ESI 
(gray squares) high-resolution analyses (collectively referred to as HR).
The left-hand panel shows a Gaussian fit to the residuals between
\afe~for our method and the HR analyses, whereas the right-hand panel displays
our determination of \afe~vs. the values for the HR analyses. The solid line is the
one-to-one line. For clarity, only a typical HR error bar is shown in
the lower-right corner.}
\label{fig:hiresafe}
\end{figure}

Figure \ref{fig:hiresafe} illustrates the results of the comparison of
our derived \afe~with the high-resolution determinations. Black dots
represent the HET stars, while the gray squares indicate the ESI
stars. The notation ``HR'' denotes the high-resolution determinations;
``SSPP'' refers to our results from the low-resolution SEGUE spectra. A
Gaussian fit to the residuals, shown in the left-hand panel, suggests a
mean offset of 0.003 dex and a standard deviation of 0.069 dex,
respectively, rather similar to those found from the comparison with the
ELODIE spectra. Using Equations (\ref{eqn:sys}) and (\ref{eqn:tot}),
we compute the total error bar on the measured \afe~and place it in the
figure. 

Examination of Figure \ref{fig:hirescor} reveals no strong correlations between our
measured residuals of \afe~with \teff, \logg, and \feh, within our derived
errors, although there may be a slight tendency toward lower \afe~determinations
at the lowest temperatures (\teff~$< 4750$~K), gravities (\logg~$< 1.8$), and
metallicities ([Fe/H] $< -2.5$), mainly for ESI stars. These behaviors 
provide information on the limitations of our technique over the parameter
space. Thus, based on present information, our method may not work for the coolest
metal-poor giant stars (\teff~$<$ 4800 K, \logg~$<$ 2.0, and [Fe/H] $< -2.5$).
These tendencies are, of course, also a natural consequence of using
low-resolution spectra to determine the \alp-abundances, as there exist few
sensitive spectral features of the \alp-elements for such low gravity,
metal-poor stars. Clearly, additional comparison stars with low temperature,
gravity, and metallicity would be useful to set the effective range of the
parameter space for the measured \afe. Nevertheless, as we can obviously
separate out the low-\alp~stars from the high-\alp~ones, we regard the parameter
coverage of the high-resolution sample as the valid range of our method.

For clarity in the plots in both Figures 7 and 8, we only show the vertical
error bars; the typical error bar from the high-resolution analysis for each
parameter is displayed in the lower-right corner of each panel. From this
comparison, we derive 0.062 dex as a typical total error in the estimated
\afe~for the HET and ESI stars, which is commensurate with that estimated from
our analysis of the ELODIE spectra. The computed $<\sigma_{\rm sys}>$ for this
sample (0.048 dex) is in good agreement with that of the ELODIE dataset (0.044
dex). The deviation scatter between this method and the high-resolution analysis
is 0.274 dex for \logg~and 0.175 dex for \feh, which are somewhat larger than
those from the ELODIE comparison. This is partly due to the fact that we adopt
\teff~from the literature to run our technique and that the ELODIE spectra have
much higher S/N.

Considering the results from both the ELODIE and high-resolution spectra
comparisons, we conclude that, in addition to reproducing the scatter of \logg~and \feh~
from the SSPP (less than 0.3 dex and 0.2 dex, respectively), we are able to
estimate \afe~with a precision on the order of 0.06 dex from SDSS/SEGUE spectra
with S/N $> 50/1$ (note that the SEGUE spectra for all of the HET and ESI stars
had high S/N, driven by the selection of brighter stars for high-resolution
observation).

\subsection{Comparison with Likely Member Stars of Globular and Open Clusters}

Stars in Galactic GCs and OCs provide a good testbed for validation of the stellar atmospheric
parameters, as in most clusters it is expected that the member stars
were born simultaneously out of well-mixed, uniform-abundance gas at the
same location in the Galaxy. Hence, member stars should exhibit very
similar elemental-abundance patterns. Note that even though there is
evidence for internal variations in the abundances of light elements in
GCs (notably C and N; Cohen et al. 2005),
since the spectral region we utilize does not contain any significant C
or N lines, our measured parameters are not affected by this variation.
This statement also holds true for Na and Al, which exhibit
star-to-star variations in the GCs (e.g., Carretta et al. 2009a, 2009b;
references therein), as the wavelength window for measuring \afe~does
not include any strong lines of these two elements. One concern in the
use the GCs as external calibrators is that there is some evidence that
there may exist a variation in [Mg/Fe] associated with Na and Al
anomalies, especially for M13 (Johnson et al. 2005). However, as we are
attempting to measure the overall content of the \alp-elements, not only
the Mg abundance, the internal dispersion of \afe~due to possible
variations of Mg abundances in the cluster will be mitigated, so we
include the GC member stars to compare with our derived \afe.

During the course of the SDSS and SEGUE surveys, we have secured photometric
and spectroscopic data for stars in the vicinity of several globular and
OCs. A subset of these clusters, M15, M13, M71, NGC~2420, M67, 
and NGC 6791, have published high-resolution spectroscopic determinations of
\alp-abundances that we use to compare with our estimates from the
low-resolution SEGUE spectra.

Following the procedures in Paper II and Paper IV, which describe how we select
likely member stars in the GCs and OCs and use them to validate the
stellar parameters delivered by the SSPP, we first select likely member
stars for each cluster. We obtain totals of 59 (M15), 217 (M13), 17 (M71), 
125 (NGC~2420), 52 (M67), and 88 (NGC 6791) respectively, of likely members 
with \afe~determination based on spectra having an average of S/N $> 20/1$ per
pixel. We then calculate the weighted average $<$\afe$>$ and its scatter
for each cluster, following Equations (\ref{eqn:mean}) and (\ref{eqn:rms}).
In this case, $x_{i}$ is the \afe~estimate and $w_{i} =
1/\sigma^{2}_{i}$, $\sigma_{i}$ being the error in \afe~of the $i$th
star. The error in the calculated mean ($<$\afe$>$) is computed from:

\begin{equation}
  \sigma^{2}_{<x>} = \frac{1}{\sum^{n}_{i=1}1/\sigma_{i}^{2}}.
\label{eqn:error}
\end{equation}

To obtain the literature value of \afe~for each cluster, we have searched for
high-resolution studies in the literature on each cluster for any available
abundances of Mg, Ti, Ca, and Si, and compute the weighted mean of \afe~from
Equation (\ref{eqn:mean}). As the error in each elemental abundance is provided in
all cases, the error in the weighted mean of \afe~is calculated by the following
equation:

\begin{equation}
  \sigma^{2} = \frac{\sum^{n}_{i=1}(w_{i}x_{i})^{2}}{({\sum^{n}_{i=1}w_{i}})^{2}},
\label{eqn:error}
\end{equation}

\noindent where $x_{i}$ is substituted by the error in the Mg, Ti, Ca, Si abundances.
The same weighting factors ($w_{i}$) are applied as for the ELODIE
and ESI spectra discussed above. Table \ref{tab:cluster}
summarizes the derived means and errors in the mean of \afe~from the
literature values and from our likely member stars for each cluster,
along with the literature and our derived estimates of [Fe/H]. Sources
for the literature values are listed in the table notes.

Figure \ref{fig:clusafe} shows, from top to bottom, the distribution of
the estimated \afe~for the selected member stars of M15, M13, M71, NGC~2420, 
M67, and NGC 6791 as a function of $(g-r)_{0}$ (left-hand panels),
\logg~(middle panels), and the average signal-to-noise ratio per pixel,
$<$S/N$>$ (right-hand panels). The dashed line is the weighted mean
from the literature, while the solid line is our weighted average of
\afe~of the likely member stars. The weighted averages ($<$\afe$>$) of
\afe~from the literature are listed at the top of each middle panel; our
weighted means ($\mu$) and scatter ($\sigma$) for each cluster are
listed at the top of each right-hand panel. The mean metallicity of each
cluster adopted from the literature is also displayed at the top of each
left-hand panel. The typical error bar, shown in the lower-left corner
of the left-hand panels, is obtained by quadratically adding the random
error of each member star and the systematic error (0.054 dex) derived
from the HET and ESI data.

Inspection of Figure \ref{fig:clusafe} indicates that our derived means
agree very well with those from the high-resolution analyses up to  
super solar metallicity ($\sim +$0.5), except
perhaps for the GC M15. As there are not many
high-resolution studies performed on this cluster, only two studies
(Sneden et al. 1997 and Sneden et al. 2000) were first
considered to derive the literature abundance of the \alp-elements.
However, we compute the mean \afe~from Sneden et al. (1997), which
reports the abundances for the four \alp-elements, because the other
study does not report Mg abundance, which has the largest contribution
on computing \afe. The apparently large discrepancy in M15 shown in
Figure \ref{fig:clusafe} is due to the high [Si/Fe] ($+$0.60) and [Ti/Fe]
($+$0.46) ratios measured by Sneden et al. (1997). The abundances of these
two elements also exhibit large star-to-star scatter in their study,
which they attribute to observational error. It will be interesting to
see how the mean values change when \alp-abundances from future
high-resolution observations become available. It is worth noting that
our measured mean \afe~($+$0.24) is much closer to that ($+0.33$) of Kirby et al.
(2008a), who derived their estimate from 44 stars in M15 using
a very similar method to ours, albeit applied to higher resolution
spectra ($R = 6000$) in a redder spectral region ($6300 - 9100$ \AA).
Their average value of \afe~is shown as the dash-dotted line in Figure \ref{fig:clusafe}.

It should also be noted that the overall metallicity of NGC 2420 derived
by Pancino et al. (2010) in Table \ref{tab:cluster} is relatively higher
(by about 0.3 to 0.4 dex) than other studies (e.g., Friel $\&$ Janes
1993; Friel et al. 2002; Anthony-Twarog et al. 2006), which are based on
medium-resolution spectra or intermediate-band photometry. Their value
is also larger by $\sim$ 0.2 dex than that of another high-resolution
study by H. Jacobson et al. (2010, in preparation), ($-0.22 \pm 0.07$), who used
nine ($R = 21,000$) spectra of giants to calculate the average
metallicity. Their overall metallicity is in fact much closer to ours
($-0.25$) in Table \ref{tab:cluster}. This cluster clearly requires more
high-resolution studies to confirm its mean metallicity. As Pancino et
al. (2010) provide the mean \afe~of the cluster with the overall [Fe/H],
we adopt it in the table.

It is noteworthy that Figure \ref{fig:clusafe} shows that even some cool metal-rich 
giants in NGC 6791 exhibit good agreement (within 2$\sigma$, considering the rms
scatter), as these stars could be considered supplementary objects to the
high-resolution sample (which lacks cool metal-rich stars). Also notice that 
although our grid of synthetic spectra reaches up to [Fe/H] = $+$0.4, 
Table \ref{tab:cluster} lists $+$0.428 as the overall metallicity of the cluster.
This arises from applying the correction function for metallicity to 
the [Fe/H] determined by this method, as explained in Section 2.3. 

Further inspection of Figure \ref{fig:clusafe} reveals that
the uncertainty in the measured \afe~increases for more metal-deficient
stars, as expected due to the overall weakness of the lines involved in
its estimation. However, no obvious covariance in the reported \afe~with
respect to $(g-r)_{0}$, \logg, and $<$S/N$>$ is noticed, although the
random scatter around the mean increases at lower S/N for
high-gravity blue stars, especially in metal-poor cases (M13 and
M15). This emphasizes the difficulty of obtaining reliable \afe~for
faint, metal-poor main-sequence turnoff and dwarf stars.

\section{Reliability of \afe~Estimation from SDSS/SEGUE Spectra}

\subsection{Effects of Declining Signal-to-oise Ratios}

While the two data sets (the ELODIE spectra and the SDSS/SEGUE
stars with parameters estimated from the high-resolution analysis) used
for the validation of our technique have very high S/N ($>$ 100/1 for
the ELODIE spectra, much higher than that when smoothed to the
SDSS resolution, and $>$ 50/1 for the SDSS/SEGUE stars with
high-resolution spectra), the low-resolution SDSS/SEGUE spectra cover a
wide range of S/N. Thus, it is desirable to check the impact that
lower S/N for a given spectrum has on estimation of \afe. Following
the prescriptions described in Paper I, we perform noise-injection
experiments on the ELODIE spectra\footnote[15]{The noise-added
spectra, and more detailed information on noise models can be found at
ftp://hebe.as.utexas.edu/pub/callende/sdssim/.} and the SDSS/SEGUE
spectra of the stars with high-resolution observations. 
Among the SDSS/SEGUE stars with high- and medium-resolution spectroscopy
available, only those with stellar parameters determined from analysis of the
HET data had noise-added spectra available, so we just consider this subset of
the ELODIE and the SDSS/SEGUE spectra in this experiment.

Table \ref{tab:noise} summarizes the results. The mean offset ($\Delta$)
and the standard deviation ($\sigma$) for each parameter are derived
from a Gaussian fit to the residuals between our results and the
external sources listed in the table. The effective temperature with the
subscript ``SSPP'' is the the adopted value from the SSPP (which was
held fixed for the determination of \afe). The listed $\sigma_{\rm tot}$
is the average total error in \afe, calculated following Equations
(\ref{eqn:sys}) and (\ref{eqn:tot}). The label ``Full'' in the column
listing S/N indicates that the parameters are derived from the spectra
prior to noise injection.

\subsubsection{{\rm HET} Comparison}

From inspection of Table \ref{tab:noise}, the effective temperature from
the SSPP that we fix to predict \afe~exhibits a systematic offset of
about 130~K, with a scatter of 180~K, from high to low S/N; we do not
correct for this offset when estimating \afe. This offset has a small dependence
on \teff, but not on \logg, nor \feh. If we add the ESI data, the offset becomes
about 140 K, as seen in Table \ref{tab:error}, and depends only very slightly on
temperature below \teff~$<$ 5000 K, as the ESI spectra mostly contain cool
stars. Nonetheless, this temperature offset does not grossly impact on our
derived \logg, \feh, and \afe, as can be deduced from Table \ref{tab:error}.
More discussion on this point is given below.

One can also notice an offset of about 0.1 dex in \logg, relative to the SSPP values,
decreasing as the S/N decreases, albeit with larger scatter. For \feh,
the offset generally increases, with larger scatter, as the quality of
the spectrum decreases. We see the same trend for \afe~in Table \ref{tab:error}.
Although there indeed exist small systematic offsets in \teff, \logg,
and \feh~(of slightly different size) at different values of S/N,
these offsets do not appear to impact estimates of \afe; only small
offsets ($< 0.01$ dex), with a relatively small scatter ($< 0.1$ dex)
are found for our \afe~determinations down to S/N = 20.

It is also clear that our technique reproduces \logg~and \feh~reasonably
well, as the rms scatter between the high-resolution analysis and
our results is $<$ 0.3 dex for \logg~and $<$ 0.2 dex for [Fe/H] down to
about S/N = 20.

\subsubsection{{\rm ELODIE} Comparison}

For the ELODIE stars, Table \ref{tab:noise} exhibits much smaller offsets
and scatters for all three parameters, \logg, \feh, and \afe. The small offsets
and scatters are the natural consequence of adopting \teff~from the literature,
and holding it fixed while estimating \afe, rather than using the internally
determined \teff.

Interpreting the results of the noise-injection tests from both the HET and
ELODIE samples, we can infer that at high S/N the dominant error in the total
uncertainty is the systematic error, $<\sigma_{\rm sys}>$ in Equation
(\ref{eqn:tot}), while at low S/N the random internal error,
$\sigma_{\rm SSPP}$ in Equation (\ref{eqn:tot}), is the dominant error, as the
total error starts smaller and becomes larger with declining S/N, as can be
seen in Table \ref{tab:noise}.

We also conclude from the noise-injection experiments that we are able to
estimate \afe~with a precision of $<$ 0.1 dex down to S/N = 20/1, while
reproducing \logg~and \feh~estimates within about 0.3 and 0.2 dex, respectively,
at the same S/N. We emphasize that, although the results of the ELODIE
comparisons appear better behaved, the quoted scatters from the comparisons with
the high-resolution analysis are more realistic, since we adopt the
\teff~derived by the SSPP to determine \afe, whereas the \teff~provided
in the literature was used for the ELODIE stars.

Some additional caution should be exercised in the interpretation of these
results, since relatively metal-rich stars (with [Fe/H] $> -1.4$) dominate in
both samples, as can be appreciated by inspection of Figures
\ref{fig:elodiecor} and \ref{fig:hirescor}. We might expect that the
error in the determination of \afe~varies with the metallicity of a star such
that the uncertainty of \afe~will be larger in the metal-poor stars than for the
metal-rich stars, as the metallic line strengths become weaker. Additional noise
in the spectrum of a metal-poor star will drive the uncertainty to even higher
values. Owing to the scarcity of metal-poor stars in our comparison samples, we
are not able to carry out a thorough test on the dependency of the uncertainty
in the measured \afe~with metallicity. Nevertheless, by only taking into account
stars with [Fe/H] $< -1.4$ at S/N = 25 and calculating the total error in the
derived \afe, we obtain $\sigma_{\rm tot} =$ 0.112 dex for the HET stars and
0.108 dex for the ELODIE stars. Thus, for the present, we suggest that spectra
of at least S/N = 25/1 are required to achieve total errors in the estimated
\afe~on the order of 0.1 dex for low-metallicity stars ([Fe/H] $< -1.4$). The
limiting lower metallicity over which our determinations of \afe~can be made
will surely vary with effective temperature (we expect sufficient line strengths
of the \alp-element features to allow determinations for cooler stars down to
[Fe/H] $\sim -3.0$, but perhaps not for warmer stars below [Fe/H] = $-2.5$).
Additional exploration will be necessary to be certain.

\subsection{Degeneracy Effects Among Parameters}

Within the spectral window (4500 -- 5500 \AA) chosen to estimate \afe,
the \ion{Mg}{1}$b$ and MgH lines are the dominant contributors to our estimates
of \afe. These spectral features are also sensitive to the surface
gravity and overall metallicity of a given star, as illustrated in
Figure \ref{fig:speclines}. The \alp-elements are also good indicators
of metallicity. Thus, one has to be concerned about possible
degeneracies among the derived stellar parameters and \afe. In other
words, does the application of our approach make it possible for an
\alp-enhanced giant to be interpreted as a dwarf with a low \afe~ratio?
Or can a metal-rich low-\alp~star be identified as a metal-deficient
star due to weakness of the \alp-element spectral features through
application of our techniques?

\begin{figure}
\centering
\plotone{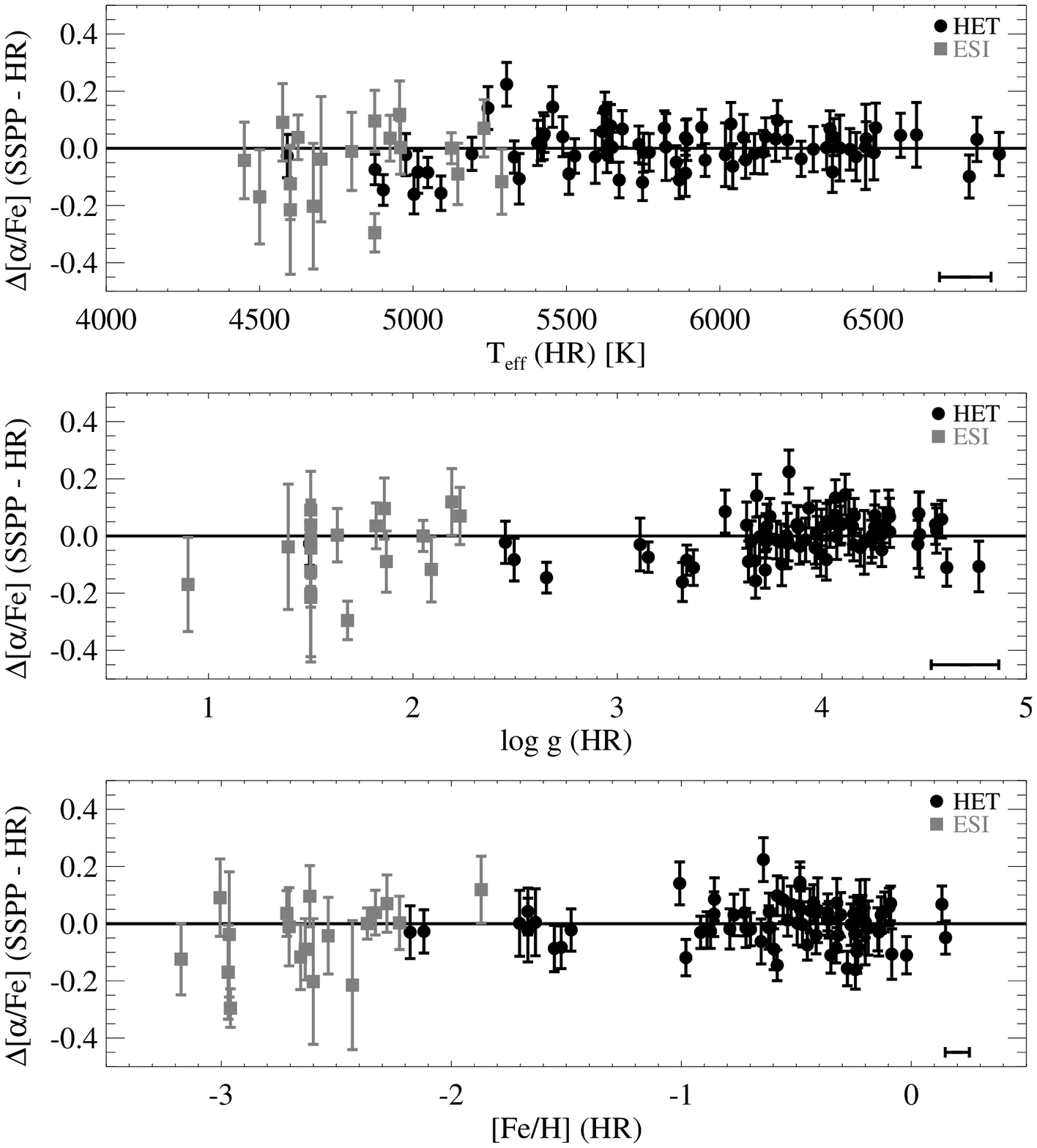}
\caption{Variations in \afe~as functions of \teff, \logg, and \feh, from
upper to lower panel, respectively, for the SDSS/SEGUE
high-resolution spectra. The black dots are the HET spectra, and 
the gray squares are the ESI spectra. Quadratically-added vertical error bars are
shown, and a typical error bar on each parameter from the
high-resolution analysis is displayed in the lower-right corner of each
panel. There appears to be little or no covariance in the determined
\afe~along with various ranges of each parameter.}
\label{fig:hirescor}
\end{figure}

\begin{figure}
\centering
\plotone{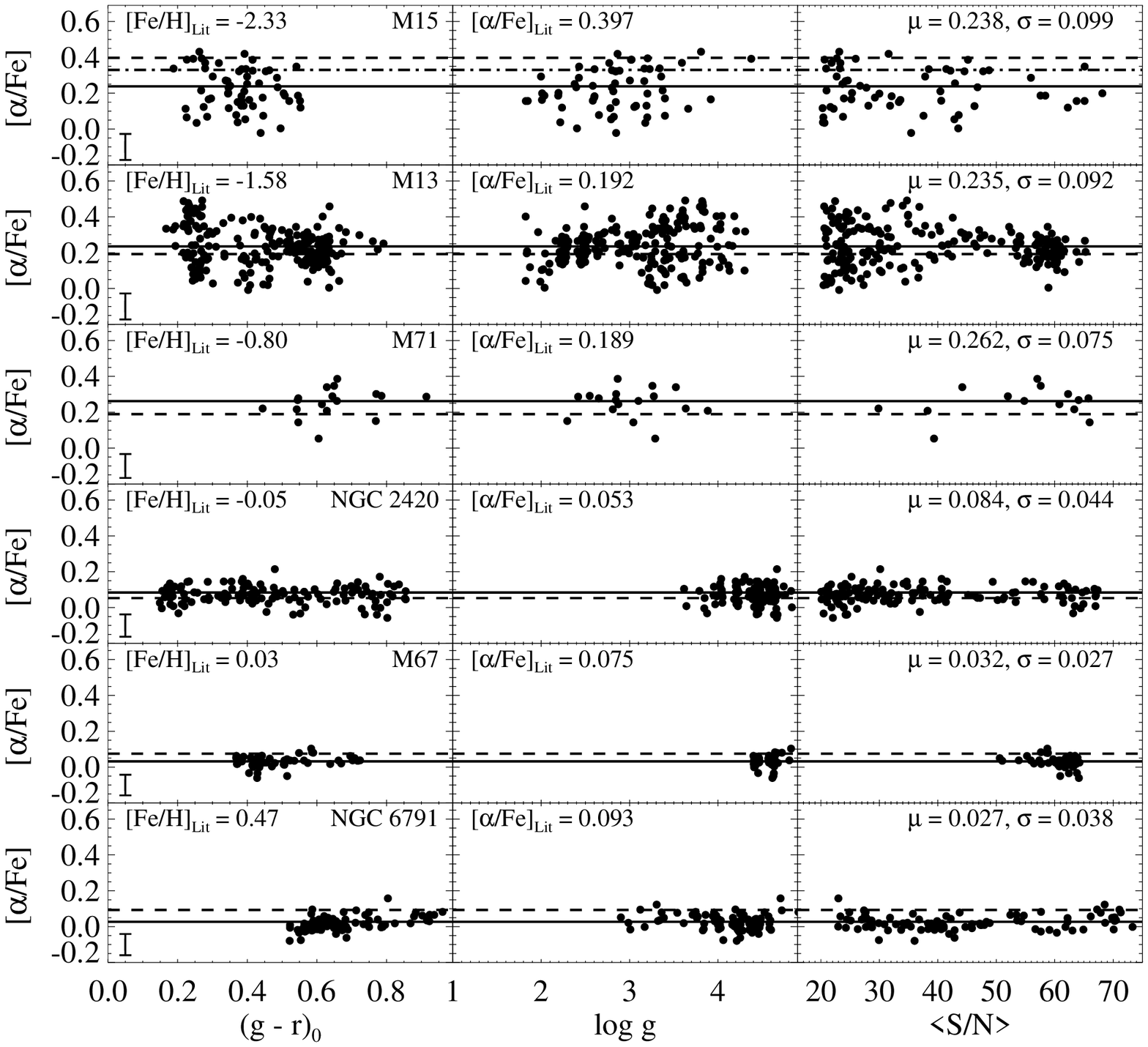}
\caption{Distribution of our measured \afe~for likely member stars of
M15, M13, M71, NGC~2420, M67, and NGC 6791 from upper to lower panels
respectively, as functions of $(g-r) _{0}$,
\logg, and $<$S/N$>$, the average signal-to-noise ratio per pixel. The solid line 
is the weighted mean of our estimated \afe~for the likely members
for each cluster, while the dashed line is a weighted average of
reported literature values for \alp-elements in each cluster (see the text).
A typical error bar in \afe~for each cluster is shown in the left-hand
panels. With the exception of M15, our \afe~estimates agree with the
averaged literature values within the star-to-star scatter among the
likely member stars. The literature determination of \afe~for M15 may
be problematic (see the text). The dash-dotted line shown for this cluster,
which appears to be in better agreement with our determination, within the
star-to-star scatter, is the average of \afe~derived by Kirby et al.
(2008a) using medium-resolution ($R=6000$) spectra of member stars in
M15. Inspection of these panels reveals little or no significant trends
in \afe~determinations for these clusters as a function of color,
surface gravity, or signal-to-noise ratio. However, the error in the
measured \afe~seems to increase as both [Fe/H] and $<$S/N$>$ decrease.}
\label{fig:clusafe}
\end{figure}

We confront this issue by considering the HET and ESI data,
which contain both dwarfs and giants covering a range of metallicities. We
examine this sample, looking for stars that are misclassified as a
result of covariance between surface gravity and the effect of
\alp-element variations on the strength of, e.g., the Mg lines. Recall
that we have already considered in Figure \ref{fig:hirescor} the
distribution of \afe~residuals between the high-resolution analysis and
our low-resolution analysis as a function of \teff, \logg, and \feh, for
different ranges of each parameter, and found that there was no
significant correlations with each parameter.

\begin{figure}
\centering
\plotone{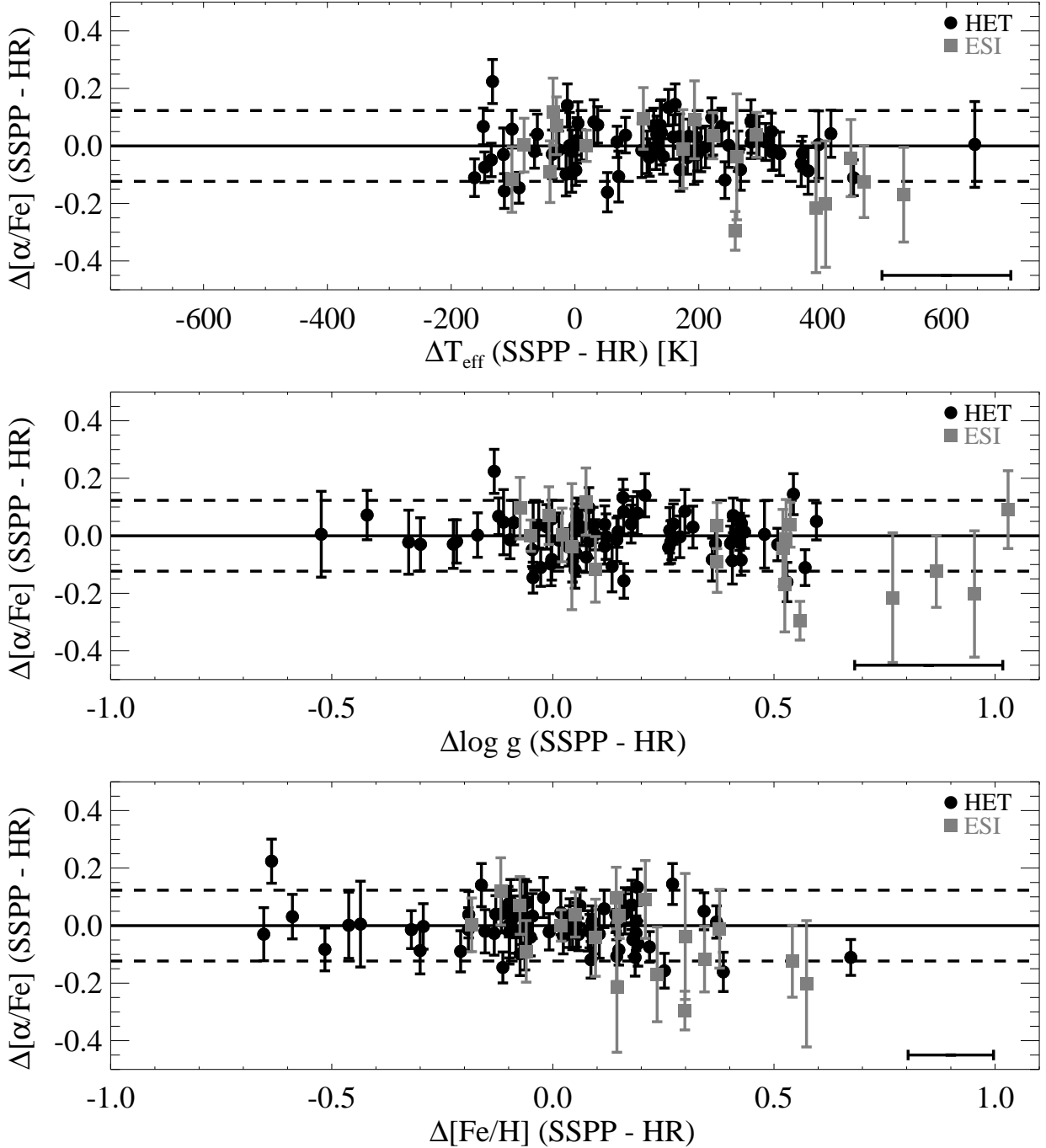}
\caption{Variations in \afe~as functions of the residuals in \teff,
\logg, and \feh, from the upper to lower panels, respectively, between
our analysis and the high-resolution analyses. The black dots are the HET spectra, and 
the gray squares are the ESI spectra. The dashed lines
are $\pm$2$\sigma_{\rm tot}$ ($\sigma_{\rm tot}$ = 0.062 dex) derived in Section 3.2. Generally, it appears
that differences in the \afe~estimates for the stars are mostly well
within $\pm$3$\sigma_{\rm tot}$ and that if our
estimates of \teff, \logg, and \feh~are higher, the determined \afe~is
slightly lower than that of the high-resolution
analysis results.}
\label{fig:hiresdiff}
\end{figure}

Figure \ref{fig:hiresdiff} shows variations in \afe~as functions of the
residuals between our analysis and the high-resolution analysis of
\teff, \logg, and \feh, from the upper to the lower panels. The dashed lines
are $\pm$2$\sigma_{\rm tot}$ ($\sigma_{\rm tot}$ = 0.062 dex) derived in Section 3.2. 
For clarity, error bars of the residuals of each parameter are not plotted
for each star; instead, an average error is drawn in the lower-right
corner of each panel. Scrutinizing the top panel of the figure reveals that,
although there are a handful of stars that deviate by more than 300~K,
differences in the \afe~estimates for these stars is mostly well within
$\pm$3$\sigma_{\rm tot}$ with the allowed error bars. Specifically, among the
stars within $\pm$300~K, 57\% are inside $\pm$1$\sigma_{\rm tot}$, 89\% for
$\pm$2$\sigma_{\rm tot}$, and 97\% for $\pm$3$\sigma_{\rm tot}$ without taking
into account the error bars. If we remove the systematic shift in
\teff~(about 140~K from Table \ref{tab:error}), almost all stars fall into the
range $\pm$300 K.

The middle panel of Figure \ref{fig:hiresdiff} indicates that the
surface gravity determined by our method mostly agrees with the
high-resolution analysis within $\pm$0.5 dex. Even though there are a
few stars with apparently large deviations ($> 0.5$ dex) in surface
gravity with respect to the high-resolution analysis, no star in this
sample would have its classification changed from a dwarf to a giant, or
vice versa. Furthermore, the great majority of stars have
\afe~differences that fall well inside $\pm$3$\sigma_{\rm tot}$;
we obtain that 61\%, 93\%, and 99\% of the stars
within $\pm$0.5 dex reside inside 1, 2, and 3$\sigma_{\rm tot}$ levels,
respectively. This suggests that even if the gravity determination is
off by over 0.5 dex, the estimated \afe~is not significantly impacted by
more than 3$\sigma_{\rm tot}$. Note that even for the very low gravity stars
(\logg~$\sim$ 1.5) seen in Figure \ref{fig:hirescor}, we consistently
obtain a good agreement with the high-resolution results.

The lower panel of Figure \ref{fig:hiresdiff} shows that the metallicity
determined by our technique is mostly consistent with the
high-resolution analysis within $\pm$0.3 dex. In this range,
69\% of the stars are inside $\pm$1$\sigma_{\rm tot}$, 89\% for $\pm$2$\sigma_{\rm tot}$,
and 97\% for $\pm$3$\sigma_{\rm tot}$. Moreover, since almost
all stars have \afe~differences that fall well inside $\pm$3$\sigma_{\rm tot}$, it
is clear that variations in the line strengths of the \alp-elements do
not result in large perturbations in the derived metallicity estimates.
This is likely the result of the application of multiple methods in the
SSPP for the determination of [Fe/H].

The confirmations from both cases above imply that we are indeed
measuring an \afe~ratio, rather than obtaining spurious estimates due to
shifts of the other parameters, such as \logg~and
\feh, that compensate for differences in the \alp-element line strengths. It is
very unlikely that our analysis would allow a high-\afe~giant to
masquerade as a low-\afe~dwarf, or vice versa. Similarly, we would not
expect a metal-rich low-\alp~star to be identified as a metal-deficient
star, due to weakness of the \alp-element spectral features, through the
application of our techniques.

However, as discussed in Section 3.2, 
there does appear to exist a weak tendency (when excluding the stars with large error bars) that if
our estimates of \teff, \logg, or \feh~are higher than the
high-resolution values, then the determined \afe~is slightly lower than
that obtained from the high-resolution analysis. The stars which display that trend 
are mostly cool metal-poor giants, and due to statistically small sample 
of such stars, it is hard to tell how much the behavior significantly affects our results. 
A supplementary set of high-resolution spectra of cool metal-poor stars may be useful 
to clarify this.  

\begin{figure}
\centering
\plotone{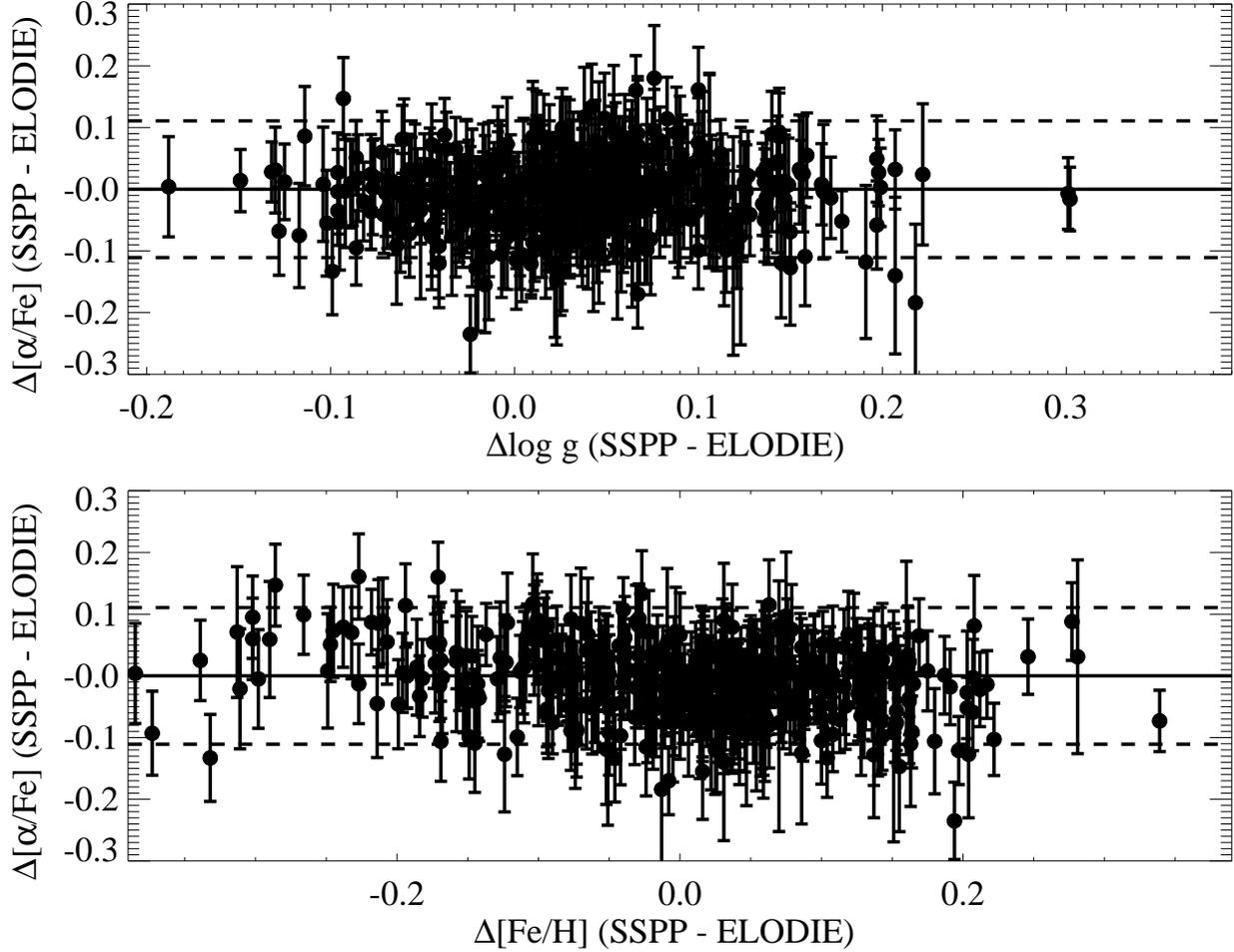}
\caption{Same as shown for Figure \ref{fig:hiresdiff}, but for the ELODIE spectra.
The dashed lines are $\pm$2$\sigma_{\rm tot}$ ($\sigma_{\rm tot}$ = 0.054 dex) derived in Section 3.1.
Note that there is no plot for \alp-abundance residuals as a function of \teff, since we
make use of that value adopted from the literature (see the text). It is noticed that differences
in the determined \afe~for the stars are mostly inside $\pm$3$\sigma_{\rm tot}$.}
\label{fig:elodiediff}
\end{figure}

Similar checks have been performed for the ELODIE spectra; Figure
\ref{fig:elodiediff} displays the results. In this figure we do not
compare against \teff, because, as mentioned previously, we adopt
\teff~from the literature (and hold it fixed during the
\afe~determination). The dashed lines are $\pm$2$\sigma_{\rm tot}$
($\sigma_{\rm tot}$ = 0.055 dex), as calculated in Section 3.2. As can be appreciated from
inspection of the upper panel, no stars exhibit gravity differences with
respect to the literature values larger than 0.5 dex; only a few stars
deviate by more than 0.3 dex in the metallicity differences from the
literature values, as seen in the lower panel. These two panels clearly
imply that there is a very tight distribution in the metallicity and the
gravity differences with respect to the literature values, with only a
few stars exhibiting \afe~differences larger than 3$\sigma_{\rm tot}$.

We conclude from these experiments that our determination of
\afe~is robust over a wide range of surface gravity ($1.5
\le $ \logg~$ \le $ 5.0) and metallicity ($-3.0 \le$ [Fe/H] $\le +0.3$), and
that it is not unduly influenced by possible degeneracies between
absorption lines with sensitivity to surface gravity and metallicity as
well as to \afe. This is confirmed by Figure \ref{fig:clusafe}; down to
\logg~$= 2$ there is no manifest trend of \afe~with the gravity estimate, 
only the random scatter around the mean of \afe~varies. The results for NGC 6791 
provide evidence that the metallicity range of our technique may 
extend to $\sim+$0.5.

\subsection{Effects of Errors in \teff~on Determination of \afe}

During the process of carrying out the minimum $\chi^{2}$ search, we have fixed
\teff~at the valued determined by the SSPP, and only allow the other three
parameters, \logg, \feh, and \afe, to be solved for simultaneously. However,
since the effective temperature estimate delivered by the SSPP itself carries
uncertainty, we need to check on how this error in \teff~propagates into
uncertainties in the determination of \afe. We perform this test (using the HET
and ESI spectra) by varying the adopted \teff~by $-$300, $-$200, $-$100, $+$100,
$+$200, and $+$300~K from the value suggested by the SSPP. Table \ref{tab:error}
summarizes the results of this experiment, and gives the derived variations in
the estimated \logg~and \feh, and \afe. All the listed mean offsets and standard
deviations in the table are derived from Gaussian fits to the residual
distributions.

There exists (as we have previously pointed out) a systematic offset of about
140~K in \teff~between the SSPP and the high-resolution analysis, even when no
perturbation is applied to the fixed \teff~when evaluating \afe, as discussed in
Section 4.1.1. Hence, this test is also useful to estimate how the temperature
offset influences our derived quantities of \logg, \feh, and \afe. According to
Table \ref{tab:error}, even if we adjust the offset by 100 K or 200 K, we do not
notice much change of the rms scatter in each parameter. At maximum, about 0.025
dex in [Fe/H] for the adjustment of $-$200 K is notable. This confirms that the
temperature offset has only a very minor impact on our measured parameters.

Table \ref{tab:error} also allows us to infer that for all three derived
parameters the mean offsets associated with different input offsets in \teff~are
rather small, and all cases are less than the derived rms variation in the
determinations of these parameters. Accordingly, it appears that within
$\pm$300~K (which is a bit larger than the typical error of 250 K of the
SSPP-determined \teff), the \afe~estimate is perturbed by less than $\pm$0.1
dex, which is smaller than 2$\sigma_{\rm tot}$. This implies that our approach
to deriving \afe~is robust against small deviations of the estimated
temperature.

\section{Summary and Conclusions}

We have presented a method for estimating \afe~from the low-resolution
($R = 2000$) SDSS/SEGUE stellar spectra, based on spectral matching
against a grid of synthetic spectra. Star-by-star comparisons with
spectra from a degraded ELODIE spectral library (with
\afe~calculated from literature values of individual \alp-element
abundances) and with the SEGUE spectra of stars with high-resolution
determinations of \afe, indicate that this approach is capable of
estimating \afe~with a precision of $<$ 0.07 dex for spectra with S/N $>
50/1$ over the parameter space \teff~= [4500, 7000] K, \logg~= [1.5,
5.0], \feh~= [--3.0, $+$0.3], and \afe~= [--0.1, $+$0.6]. According to
our noise-injection experiments, errors in the determination of
\afe~increase to $>$ 0.1 dex for S/N $< 15/1$. Thus, for application of
this approach, we recommend that spectra with a minimum S/N = 20/1 be
used. From close examination of a small sample of metal-poor stars ([Fe/H] $<
-1.4$), we have found that it is desirable to have spectra with S/N $> 25/1$ for
such stars to achieve an uncertainty in the measured
\afe~comparable to 0.1 dex. Using our methods, we obtained $\sigma_{\rm tot}$ = 0.112 dex
from the HET and ESI stars, and 0.108 dex from the ELODIE sample at this
S/N for these metal-poor stars. The question of whether this S/N limit could
be lowered will require a larger number of comparison stars at low metallicity
to be considered, an effort that is presently underway.

A validation with likely members of the Galactic GCs M15, M13, and M71 
and the OCs NGC~2420, M67, and NGC 6791 confirms that the weighted average
\afe~for member stars obtained by our measurements is in good agreement with the
weighted mean of the literature values, within the reported scatter.
The one possible exception is M15, which has a published value for the Mg
abundance, with large star-to-star scatters for [Si/Fe] and [Ti/Fe],
based on high-resolution spectroscopy. Our estimate of \afe~for this
cluster is, however, close to that of Kirby et al. (2008a), who employed
a similar technique as ours for somewhat higher resolution spectra of member 
stars in M15. The comparison with NGC 6791 implies that the valid range of the 
metallicity for our method may extend to $\sim+$0.5.

We have looked for, and failed to find, any significant covariance between
surface gravity and our derived estimates of \afe, which depend on the same
spectral features for much of their sensitivity to these parameters. We have
also found relatively small variations in \logg~($<$ 0.3 dex) and \feh~($<$ 0.2 dex)
with \afe~(itself mostly exhibiting peak-to-peak scatter of less than 3$\sigma_{\rm tot}$)
between our measurements and high-resolution analyses.

We have also checked on possible errors in our determination of \afe~due to our
choice to fix the input \teff~to the values delivered by the SSPP (which itself
can have errors up to 250~K), and find that for errors less than $\pm$300~K,
the \afe~measurement is perturbed by less than $\pm$0.1 dex,
which is smaller than 2$\sigma_{\rm tot}$.

As this method can be easily applied to other spectra that cover similar
wavelength ranges at similar resolving power, it should be a useful new
tool for investigation of the star formation and chemical enrichment history of
Galactic populations with SDSS/SEGUE, as well as for much larger stellar
samples in the future, such as will be obtained from LAMOST.

\acknowledgements{

Funding for the SDSS and SDSS-II has been provided by the Alfred P. Sloan
Foundation, the Participating Institutions, the National Science Foundation, the
U.S. Department of Energy, the National Aeronautics and Space Administration,
the Japanese Monbukagakusho, the Max Planck Society, and the Higher Education
Funding Council for England. The SDSS Web site is http://www.sdss.org/.

The SDSS is managed by the Astrophysical Research Consortium for the
Participating Institutions. The Participating Institutions are the American
Museum of Natural History, Astrophysical Institute Potsdam, University of Basel,
University of Cambridge, Case Western Reserve University, University of Chicago,
Drexel University, Fermilab, the Institute for Advanced Study, the Japan
Participation Group, Johns Hopkins University, the Joint Institute for Nuclear
Astrophysics, the Kavli Institute for Particle Astrophysics and Cosmology, the
Korean Scientist Group, the Chinese Academy of Sciences (LAMOST), Los Alamos
National Laboratory, the Max-Planck-Institute for Astronomy (MPIA), the
Max-Planck-Institute for Astrophysics (MPA), New Mexico State University, Ohio
State University, University of Pittsburgh, University of Portsmouth, Princeton
University, the United States Naval Observatory, and the University of
Washington.

The Hobby Eberly Telescope (HET) is a joint project of the University of Texas
at Austin, the Pennsylvania State University, Stanford University,
Ludwig-Maximilians-Universit\"at M\"unchen, and Georg-August-Universit\"at
G\"ottingen. The HET is named in honor of its principal benefactors, William P.
Hobby and Robert E. Eberly. Some of the data presented herein were obtained at
the W.M. Keck Observatory, which is operated as a scientific partnership among
the California Institute of Technology, the University of California and the
National Aeronautics and Space Administration. The Observatory was made possible
by the generous financial support of the W.M. Keck Foundation. The authors wish
to recognize and acknowledge the very significant cultural role and reverence
that the summit of Mauna Kea has always had within the indigenous Hawaiian
community. We are most fortunate to have the opportunity to conduct observations
from this mountain.

Y.S.L. and T.C.B. acknowledge partial funding of this work from grants
PHY 02-16783 and PHY 08-22648: Physics Frontier Center/Joint Institute
for Nuclear Astrophysics (JINA), awarded by the U.S. National Science
Foundation. H.L.M. acknowledges support from AST-0607518. D.K.L. acknowledges 
the support of the National Science Foundation through the
NSF Astronomy and Astrophysics Postdoctoral Fellowship under award
AST-0802292. J.A.J. acknowledges support from NSF grant AST-0607482. We 
are grateful for the insightful comments of the anonymous referee, which 
contributed to numerous improvements in our paper.}

\clearpage

\begin{deluxetable}{cc|ccccc|cccccc}
\tablewidth{0in}
\tabletypesize{\scriptsize}
\renewcommand{\tabcolsep}{2pt} \tablecaption{List of Parameters for Selected ELODIE Spectra}
\tablehead{\multicolumn{2}{c}{ID} & \multicolumn{5}{c}{ELODIE} & \multicolumn{6}{c}{SSPP} \\
\cline{1-2} \cline{3-7} \cline{8-13}
\colhead{} & \colhead{ELODIE}       & \colhead{\teff} & \colhead{\logg}                  & \colhead{\feh}  & \colhead{\afe} & \colhead{$\sigma_{\rm [\alpha/Fe]}$} &
\colhead{\logg}    & \colhead{$\sigma_{\log g}$}  & \colhead{\feh}  & \colhead{$\sigma_{\rm [Fe/H]}$}  & \colhead{\afe}  & \colhead{$\sigma_{\rm [\alpha/Fe]}$} \\
\colhead{}         & \colhead{}                    & \colhead{(K)}  & \colhead{}                  & \colhead{} & \colhead{}   & \colhead{(dex)} &
\colhead{} & \colhead{(dex)} & \colhead{}  & \colhead{(dex)}  & \colhead{}       & \colhead{(dex)}}
\startdata
    HD000245 & 00001 &  5433 &   3.50 &   $-$0.76 &   $+$0.34 &   0.02 &   3.801 &   0.015 &   $-$0.757 &   0.036 &   $+$0.333 &   0.031 \\
    HD000400 & 00003 &  6146 &   4.09 &   $-$0.28 &   $+$0.11 &   0.01 &   4.060 &   0.022 &   $-$0.290 &   0.045 &   $+$0.106 &   0.029 \\
    HD000693 & 00004 &  6156 &   4.13 &   $-$0.42 &   $+$0.10 &   0.03 &   4.156 &   0.032 &   $-$0.443 &   0.030 &   $+$0.079 &   0.070 \\
    HD001835 & 00006 &  5777 &   4.45 &   $+$0.17 &   $+$0.01 &   0.03 &   4.399 &   0.010 &   $+$0.301 &   0.037 &   $+$0.019 &   0.023 \\
    HD003268 & 00010 &  6130 &   4.01 &   $-$0.24 &   $+$0.12 &   0.02 &   3.991 &   0.018 &   $-$0.177 &   0.029 &   $+$0.040 &   0.049 \\
    HD003567 & 00012 &  5991 &   3.96 &   $-$1.25 &   $+$0.21 &   0.03 &   3.930 &   0.023 &   $-$1.434 &   0.042 &   $+$0.177 &   0.037 \\
    HD003628 & 00013 &  5701 &   4.06 &   $-$0.19 &   $+$0.18 &   0.07 &   4.046 &   0.017 &   $-$0.138 &   0.040 &   $+$0.125 &   0.025 \\
    HD004307 & 00018 &  5806 &   4.04 &   $-$0.25 &   $+$0.03 &   0.03 &   4.039 &   0.021 &   $-$0.218 &   0.032 &   $+$0.021 &   0.027 \\
    HD004614 & 00020 &  5890 &   4.40 &   $-$0.28 &   $+$0.12 &   0.04 &   4.372 &   0.009 &   $-$0.193 &   0.021 &   $+$0.063 &   0.035 \\
    HD006582 & 00028 &  5313 &   4.40 &   $-$0.83 &   $+$0.36 &   0.03 &   4.558 &   0.023 &   $-$0.975 &   0.055 &   $+$0.251 &   0.059 \\
    HD006582 & 00029 &  5313 &   4.40 &   $-$0.83 &   $+$0.36 &   0.03 &   4.537 &   0.007 &   $-$0.869 &   0.057 &   $+$0.396 &   0.016 \\
    HD006920 & 00032 &  5808 &   3.47 &   $-$0.11 &   $+$0.08 &   0.07 &   3.485 &   0.020 &   $-$0.220 &   0.044 &   $+$0.094 &   0.035 \\
    HD009562 & 00034 &  5821 &   3.98 &   $+$0.15 &   $+$0.08 &   0.07 &   3.983 &   0.012 &   $+$0.290 &   0.071 &   $+$0.010 &   0.040 \\
    HD009562 & 00035 &  5821 &   3.98 &   $+$0.15 &   $+$0.08 &   0.07 &   4.000 &   0.036 &   $+$0.278 &   0.041 &   $+$0.015 &   0.049 \\
    HD010145 & 00038 &  5673 &   4.40 &   $-$0.01 &   $+$0.12 &   0.02 &   4.482 &   0.015 &   $-$0.155 &   0.048 &   $+$0.079 &   0.016 \\
    HD010307 & 00039 &  5862 &   4.27 &   $+$0.00 &   $+$0.09 &   0.05 &   4.249 &   0.016 &   $+$0.041 &   0.023 &   $+$0.059 &   0.050 \\
    HD010476 & 00041 &  5206 &   4.34 &   $-$0.03 &   $+$0.07 &   0.04 &   4.452 &   0.015 &   $+$0.078 &   0.068 &   $+$0.052 &   0.043 \\
    HD010700 & 00042 &  5313 &   4.32 &   $-$0.52 &   $+$0.32 &   0.07 &   4.475 &   0.014 &   $-$0.676 &   0.024 &   $+$0.352 &   0.030 \\
    HD010780 & 00043 &  5369 &   4.31 &   $+$0.05 &   $+$0.06 &   0.06 &   4.429 &   0.013 &   $+$0.080 &   0.034 &   $+$0.011 &   0.024 \\
    HD013783 & 00050 &  5451 &   4.14 &   $-$0.67 &   $+$0.39 &   0.04 &   4.276 &   0.015 &   $-$0.690 &   0.047 &   $+$0.340 &   0.031 \\
    HD013974 & 00051 &  5597 &   3.92 &   $-$0.41 &   $+$0.22 &   0.11 &   4.111 &   0.019 &   $-$0.461 &   0.051 &   $+$0.102 &   0.036 \\
    HD013974 & 00052 &  5597 &   3.92 &   $-$0.41 &   $+$0.22 &   0.11 &   4.127 &   0.015 &   $-$0.379 &   0.048 &   $+$0.080 &   0.045
\enddata
\tablecomments{Note that if only [Mg/Fe] is available from the published studies, no
error in \afe~is reported, and that there are multiple observations made for some stars. There are 293 unique stars out of
425 spectra. \\
This table is available in its entirety in machine-readable and Virtual Observatory (VO) forms in the online journal. A
portion is shown here for guidance regarding its form and content.}
\label{tab:elodie}
\end{deluxetable}

\clearpage

\begin{deluxetable}{cc|cccccccc|ccccccccc}
\tablewidth{0in} \rotate
\tabletypesize{\scriptsize}
\renewcommand{\tabcolsep}{2pt} \tablecaption{List of Parameters for HET and ESI Spectra}
\tablehead{\multicolumn{2}{c}{} & \multicolumn{8}{c}{High-resolution Analysis} & \multicolumn{8}{c}{SSPP} & \multicolumn{1}{c}{} \\
\cline{3-10} \cline{11-18}
\colhead{Plate-MJD-Fiber} & \colhead{SDSS Designation}        & \colhead{\teff} & \colhead{$\sigma_{T_{\rm eff}}$} &
\colhead{\logg}            & \colhead{$\sigma_{\log g}$}      & \colhead{\feh}  & \colhead{$\sigma_{\rm [Fe/H]}$}  &
\colhead{\afe}             & \colhead{$\sigma_{\rm [\alpha/Fe]}$} & \colhead{\teff} & \colhead{$\sigma_{T_{\rm eff}}$} &
\colhead{\logg}            & \colhead{$\sigma_{\log g}$}      & \colhead{\feh}  & \colhead{$\sigma_{\rm [Fe/H]}$}  &
\colhead{\afe}             & \colhead{$\sigma_{\rm [\alpha/Fe]}$} & \colhead{Ref.} \\
\colhead{}                 & \colhead{}                          & \colhead{(K)} & \colhead{(K)} &
\colhead{}            & \colhead{(dex)}      & \colhead{}  & \colhead{(dex)}  &
\colhead{}             & \colhead{(dex)} & \colhead{(K)} & \colhead{(K)} &
\colhead{}            & \colhead{(dex)}      & \colhead{}  & \colhead{(dex)}  &
\colhead{}             & \colhead{(dex)} & \colhead{}}
\startdata

  0353-51703-605 & SDSS J171652.50$+$603926.9 &   5672 &  44 & 3.372 & 0.068 &  $-$0.350 & 0.032 &  $+$0.082 & 0.029 &   6122 &  46 & 3.942 & 0.022 &  $+$0.324 & 0.112 &  $+$0.029 & 0.027 &   HET \\
  0380-51792-236 & SDSS J225801.77$+$000643.1 &   6838 &  90 & 4.258 & 0.090 &  $-$0.307 & 0.044 &  $+$0.180 & 0.050 &   6996 &  50 & 4.244 & 0.031 &  $-$0.896 & 0.161 &  $+$0.211 & 0.035 &   HET \\
  0396-51816-605 & SDSS J010746.51$+$011402.6 &   5346 &  11 & 4.768 & 0.088 &  $-$0.085 & 0.018 &  $+$0.077 & 0.046 &   5416 & 240 & 4.902 & 0.017 &  $+$0.060 & 0.083 &  $+$0.029 & 0.058 &   HET \\
  0401-51788-407 & SDSS J014149.73$+$010720.2 &   4876 &  23 & 3.151 & 0.052 &  $-$0.452 & 0.021 &  $+$0.351 & 0.018 &   4730 & 100 & 3.227 & 0.024 &  $-$0.233 & 0.107 &  $+$0.277 & 0.015 &   HET \\
  0401-51788-410 & SDSS J014215.40$+$011400.6 &   5417 &  40 & 3.977 & 0.067 &  $-$0.538 & 0.032 &  $+$0.334 & 0.022 &   5701 &  30 & 4.411 & 0.014 &  $-$0.168 & 0.044 &  $+$0.347 & 0.020 &   HET \\
  0409-51871-449 & SDSS J024740.30$+$011144.9 &   5868 &  18 & 4.611 & 0.123 &  $-$0.021 & 0.027 &  $+$0.132 & 0.044 &   5705 &  29 & 4.584 & 0.006 &  $+$0.165 & 0.009 &  $+$0.022 & 0.006 &   HET \\
  0409-51871-562 & SDSS J025046.89$+$010910.8 &   5527 &  43 & 3.836 & 0.067 &  $-$0.873 & 0.035 &  $+$0.355 & 0.026 &   5894 &  19 & 4.238 & 0.027 &  $-$0.685 & 0.078 &  $+$0.328 & 0.026 &   HET \\
  0421-51821-439 & SDSS J005826.06$+$150153.6 &   5003 &  26 & 3.318 & 0.053 &  $-$0.243 & 0.021 &  $+$0.151 & 0.017 &   5055 &  86 & 3.847 & 0.020 &  $+$0.142 & 0.104 &  $+$0.010 & 0.046 &   HET \\
  0434-51885-133 & SDSS J074705.19$+$414452.1 &   5048 &  28 & 3.339 & 0.056 &  $-$0.327 & 0.023 &  $+$0.196 & 0.018 &   5049 &  81 & 3.765 & 0.016 &  $-$0.178 & 0.055 &  $+$0.112 & 0.013 &   HET \\
  0441-51868-497 & SDSS J082253.87$+$471741.9 &   6042 &  62 & 3.994 & 0.074 &  $-$0.653 & 0.039 &  $+$0.153 & 0.046 &   6407 &  38 & 4.047 & 0.017 &  $-$0.460 & 0.047 &  $+$0.091 & 0.042 &   HET \\
  0598-52316-443 & SDSS J115520.82$+$654309.8 &   5632 &  49 & 3.800 & 0.071 &  $-$0.888 & 0.038 &  $+$0.284 & 0.030 &   5950 &  33 & 4.221 & 0.016 &  $-$0.896 & 0.060 &  $+$0.262 & 0.028 &   HET \\
  0604-52079-572 & SDSS J134901.58$+$640924.7 &   5819 &  41 & 4.159 & 0.060 &  $-$0.090 & 0.028 &  $+$0.047 & 0.022 &   5946 &  81 & 4.567 & 0.022 &  $+$0.088 & 0.079 &  $+$0.118 & 0.030 &   HET \\
  0732-52221-345 & SDSS J213818.92$+$123547.8 &   5192 &  35 & 3.728 & 0.064 &  $-$0.699 & 0.031 &  $+$0.391 & 0.018 &   5126 &  44 & 3.811 & 0.021 &  $-$0.615 & 0.085 &  $+$0.372 & 0.028 &   HET \\
  0740-52263-364 & SDSS J224610.22$+$145156.7 &   5455 &  26 & 4.114 & 0.097 &  $-$0.484 & 0.030 &  $+$0.037 & 0.039 &   5616 &  52 & 4.658 & 0.017 &  $-$0.213 & 0.040 &  $+$0.182 & 0.036 &   HET \\
  0744-52251-179 & SDSS J231427.16$+$134821.9 &   5091 &  27 & 3.676 & 0.051 &  $-$0.278 & 0.021 &  $+$0.236 & 0.016 &   4976 &  71 & 3.837 & 0.008 &  $-$0.026 & 0.051 &  $+$0.079 & 0.033 &   HET \\
  0747-52234-136 & SDSS J233720.37$+$140953.8 &   6444 &  49 & 4.470 & 0.111 &  $-$0.141 & 0.033 &  $+$0.059 & 0.067 &   6407 &  45 & 4.245 & 0.017 &  $-$0.036 & 0.020 &  $+$0.030 & 0.019 &   HET \\
  0747-52234-212 & SDSS J233611.86$+$140923.9 &   5941 &  51 & 4.064 & 0.068 &  $-$0.425 & 0.033 &  $+$0.097 & 0.029 &   5977 &  32 & 4.244 & 0.030 &  $-$0.526 & 0.049 &  $+$0.170 & 0.029 &   HET \\
  0781-52373-015 & SDSS J124826.99$+$614358.8 &   5952 &  48 & 3.967 & 0.066 &  $-$0.318 & 0.031 &  $+$0.113 & 0.029 &   6133 &  26 & 4.229 & 0.007 &  $-$0.225 & 0.034 &  $+$0.072 & 0.013 &   HET \\
  0812-52352-578 & SDSS J155509.18$+$495003.3 &   5638 &  38 & 4.332 & 0.058 &  $-$0.365 & 0.028 &  $+$0.141 & 0.021 &   5705 &  34 & 4.480 & 0.021 &  $-$0.276 & 0.078 &  $+$0.155 & 0.015 &   HET \\
  0835-52326-601 & SDSS J111901.08$+$054319.4 &   5644 &  15 & 4.474 & 0.127 &  $-$0.223 & 0.028 &  $+$0.034 & 0.050 &   5648 &  38 & 4.665 & 0.019 &  $-$0.282 & 0.043 &  $+$0.113 & 0.028 &   HET \\
  0888-52339-599 & SDSS J074151.21$+$275319.8 &   6475 &  42 & 4.478 & 0.159 &  $-$0.200 & 0.036 &  $+$0.127 & 0.138 &   7120 &  56 & 3.954 & 0.020 &  $-$0.635 & 0.104 &  $+$0.132 & 0.030 &   HET \\
  0889-52663-204 & SDSS J074300.91$+$285106.6 &   6348 &  72 & 4.247 & 0.080 &  $-$0.496 & 0.041 &  $+$0.100 & 0.049 &   6596 &  32 & 4.077 & 0.023 &  $-$0.443 & 0.029 &  $+$0.103 & 0.036 &   HET 
\enddata
\tablecomments{Note that \teff~in the SSPP columns is delivered by the SSPP, whereas \logg, \feh, and \afe~come from the method described in
this paper. The gravity and metallicity of the high-resolution analysis are estimated from the technique addressed in Section 3.2. \\
This table is available in its entirety in machine-readable and Virtual Observatory (VO) forms in the online journal. A
portion is shown here for guidance regarding its form and content.}
\label{tab:hires}
\end{deluxetable}

\clearpage

\begin{deluxetable}{ccc|ccccc}
\tablewidth{0in}
\tabletypesize{\scriptsize}
\renewcommand{\tabcolsep}{2pt} \tablecaption{Weighted Average \feh~and \afe~for Likely Cluster Members}
\tablehead{\multicolumn{1}{c}{} & \multicolumn{2}{c}{Literature} & \multicolumn{5}{c}{SSPP} \\
\cline{2-3} \cline{4-8}
\colhead{Cluster} & \colhead{$<\rm [Fe/H]>$} & \colhead{$<$\afe$>$} &
\colhead{$N_{\rm mem}$} & \colhead{$<\rm [Fe/H]>$} & \colhead{$\sigma$(\feh)} & \colhead{$<$\afe$>$} & \colhead{$\sigma$(\afe)} \\
\colhead{} & \colhead{} & \colhead{} & \colhead{} & \colhead{} & \colhead{(dex)} & \colhead{} & \colhead{(dex)}}
\startdata
     M15 & $-$2.330  $\pm$   0.020 &    0.397  $\pm$   0.149 &    59 &  $-$2.170  $\pm$  0.010 &   0.282 &   $+$0.238  $\pm$  0.006 &   0.099 \\
     M13 & $-$1.580  $\pm$   0.040 &    0.192  $\pm$   0.017 &   217 &  $-$1.567  $\pm$  0.005 &   0.189 &   $+$0.235  $\pm$  0.003 &   0.092 \\
     M71 & $-$0.800  $\pm$   0.020 &    0.189  $\pm$   0.189 &    17 &  $-$0.755  $\pm$  0.014 &   0.081 &   $+$0.262  $\pm$  0.010 &   0.075 \\
NGC 2420 & $-$0.050  $\pm$   0.020 &    0.053  $\pm$   0.029 &   125 &  $-$0.300  $\pm$  0.006 &   0.136 &   $+$0.084  $\pm$  0.003 &   0.044 \\
     M67 & $+$0.030  $\pm$   0.010 &    0.075  $\pm$   0.012 &    52 &  $+$0.068  $\pm$  0.007 &   0.065 &   $+$0.032  $\pm$  0.004 &   0.027 \\
NGC 6791 & $+$0.470  $\pm$   0.070 &    0.093  $\pm$   0.093 &    88 &  $+$0.428  $\pm$  0.005 &   0.084 &   $+$0.027  $\pm$  0.003 &   0.038 
\enddata
\tablecomments{The literature value of $<$\afe$>$ is calculated from the following
references---M15: Sneden et al. (1997); M13: Sneden et al. (2004)
and Cohen $\&$ Mel$\acute{\rm e}$ndez (2005); M71: Boesgaard et al. (2005); NGC 2420: Pancino et al. (2009); 
M67: Pancino et al. (2010); Yong et al. (2005); Randich et al. (2006); NGC 6791: Carretta et al. (2007). 
The adopted literature values of \feh~for M15 and M13 come from Carretta et al. (2009c), Boesgaard et al. (2005) for M71,
Pancino et al. (2010) for NGC 2420 and M67, and Carretta et al. (2007) for NGC 6791. $N_{\rm mem}$ is the total number 
of the member stars considered in averaging. See the text for methods used to compute the mean and variance from both the literature values and our
derived values for each cluster.}
\label{tab:cluster}
\end{deluxetable}

\clearpage

\begin{deluxetable}{cccccccccc|cccccccc}
\tablewidth{0in}
\rotate
\tabletypesize{\scriptsize}
\renewcommand{\tabcolsep}{2pt} \tablecaption{Variation on Derived Parameters with
S/N}
\tablehead{\multicolumn{10}{c}{HET} & \multicolumn{8}{c}{ELODIE} \\
\cline{1-10} \cline{11-18}
\multicolumn{1}{c}{} & \multicolumn{2}{c}{\teff$_{\rm SSPP}$} & \multicolumn{2}{c}{\logg} & \multicolumn{2}{c}{\feh} &
\multicolumn{3}{c}{\afe} & \multicolumn{1}{c}{} & \multicolumn{2}{c}{\logg} & \multicolumn{2}{c}{\feh} & \multicolumn{3}{c}{\afe} \\
\cline{2-3} \cline{4-5} \cline{6-7} \cline{8-10} \cline{12-13} \cline{14-15} \cline{16-18}
\colhead{S/N} & \colhead{$\Delta$} & \colhead{$\sigma$} & \colhead{$\Delta$} & \colhead{$\sigma$} & \colhead{$\Delta$} & \colhead{$\sigma$} & \colhead{$\Delta$} & \colhead{$\sigma$} & \colhead{$\sigma_{\rm tot}$} & \colhead{S/N} & \colhead{$\Delta$} & \colhead{$\sigma$} & \colhead{$\Delta$} & \colhead{$\sigma$} & \colhead{$\Delta$} & \colhead{$\sigma$} & \colhead{$\sigma_{\rm tot}$} \\
\colhead{} & \colhead{(K)} & \colhead{(K)} & \colhead{(dex)} & \colhead{(dex)} & \colhead{(dex)} & \colhead{(dex)} & \colhead{(dex)} & \colhead{(dex)} & \colhead{(dex)} & \colhead{} & \colhead{(dex)} & \colhead{(dex)} & \colhead{(dex)} & \colhead{(dex)} & \colhead{(dex)} & \colhead{(dex)} & \colhead{(dex)}}
\startdata
 10 &  $+$155 & 190 &  $+$0.061 &  0.393 &  $+$0.144 &  0.298 & $+$0.004 &  0.129 &  0.122 &       6.3 &   $+$0.019 &    0.099 & $+$0.758 &    0.156 & $-$0.004 &    0.133 &   0.125 \\ 
 15 &  $+$127 & 182 &  $+$0.083 &  0.325 &  $+$0.199 &  0.193 & $-$0.003 &  0.106 &  0.099 &      12.5 &   $+$0.007 &    0.106 & $+$0.218 &    0.194 & $-$0.015 &    0.106 &   0.098 \\
 20 &  $+$122 & 181 &  $+$0.088 &  0.297 &  $+$0.185 &  0.191 & $-$0.003 &  0.095 &  0.086 &      25.0 &   $+$0.020 &    0.100 & $+$0.066 &    0.126 & $-$0.004 &    0.076 &   0.068 \\
 25 &  $+$132 & 183 &  $+$0.089 &  0.266 &  $+$0.140 &  0.172 & $-$0.005 &  0.091 &  0.083 &      50.0 &   $+$0.023 &    0.096 & $+$0.004 &    0.119 & $-$0.001 &    0.067 &   0.059 \\
 30 &  $+$132 & 182 &  $+$0.094 &  0.257 &  $+$0.104 &  0.172 & $-$0.006 &  0.088 &  0.079 &      Full &   $+$0.035 &    0.092 & $+$0.019 &    0.122 & $-$0.010 &    0.062 &   0.055 \\
 40 &  $+$134 & 175 &  $+$0.112 &  0.251 &  $+$0.067 &  0.176 & $-$0.003 &  0.088 &  0.078 &   \nodata & \nodata &  \nodata &  \nodata &  \nodata &  \nodata & \nodata  & \nodata \\
 50 &  $+$149 & 174 &  $+$0.124 &  0.247 &  $+$0.059 &  0.177 & $-$0.001 &  0.085 &  0.075 &   \nodata & \nodata &  \nodata &  \nodata &  \nodata &  \nodata & \nodata  & \nodata \\
Full&  $+$131 & 180 &  $+$0.124 &  0.250 &  $+$0.019 &  0.171 & $+$0.005 &  0.065 &  0.057 &   \nodata & \nodata &  \nodata &  \nodata &  \nodata &  \nodata & \nodata  & \nodata \\  
\enddata
\tablecomments{\teff$_{\rm SSPP}$ is the adopted temperature from the SSPP. $\sigma_{\rm tot}$ is the total error in
the estimated \afe, computed by Equations (4) and (5).}
\label{tab:noise}
\end{deluxetable}

\clearpage

\begin{deluxetable}{cccccccccc}
\tablewidth{0in}
\tabletypesize{\scriptsize}
\renewcommand{\tabcolsep}{2pt} \tablecaption{Effects of Errors in \teff~on Determination of \afe}
\tablehead{\colhead{\teff} & \multicolumn{2}{c}{\teff} & \multicolumn{2}{c}{\logg} & \multicolumn{2}{c}{\feh} & \multicolumn{3}{c}{\afe} \\
\cline{2-3}  \cline{4-5}  \cline{6-7}  \cline{8-10}
\colhead{Error} & \colhead{$\Delta$} & \colhead{$\sigma$} & \colhead{$\Delta$} & \colhead{$\sigma$} & \colhead{$\Delta$} & \colhead{$\sigma$} & \colhead{$\Delta$} & \colhead{$\sigma$}  & \colhead{$\sigma_{\rm tot}$}\\
\colhead{(K)} & \colhead{(K)} & \colhead{(K)} & \colhead{(dex)} & \colhead{(dex)} & \colhead{(dex)} & \colhead{(dex)} & \colhead{(dex)} & \colhead{(dex)} & \colhead{(dex)}}
\startdata
 $-$300  & \nodata & \nodata  & $+$0.109 &   0.266 &  $+$0.047 &   0.210 &  $+$0.000 &   0.087 &   0.079  \\
 $-$200  & \nodata & \nodata  & $+$0.131 &   0.271 &  $+$0.058 &   0.199 &  $-$0.001 &   0.080 &   0.072  \\
 $-$100  & \nodata & \nodata  & $+$0.124 &   0.271 &  $+$0.047 &   0.208 &  $-$0.007 &   0.071 &   0.063  \\
 $+$  0  & 143\tablenotemark{1} & 194\tablenotemark{1}      & $+$0.137 &   0.274 &  $+$0.042 &   0.175 &  $+$0.003 &   0.069 &   0.062  \\
 $+$100  & \nodata & \nodata  & $+$0.144 &   0.261 &  $+$0.044 &   0.189 &  $+$0.002 &   0.073 &   0.063  \\
 $+$200  & \nodata & \nodata  & $+$0.179 &   0.269 &  $+$0.069 &   0.190 &  $+$0.008 &   0.094 &   0.086  \\
 $+$300  & \nodata & \nodata  & $+$0.183 &   0.253 &  $+$0.073 &   0.194 &  $+$0.023 &   0.111 &   0.099  \\
\enddata
\tablenotetext{1}{Gaussian mean and scatter between the SSPP and the high-resolution analysis.}
\label{tab:error}
\end{deluxetable}

\clearpage

\clearpage

\clearpage

\clearpage

\clearpage

\clearpage

\clearpage

\clearpage

\clearpage

\clearpage

\end{document}